\documentclass[review]{elsarticle}

\usepackage{lineno,hyperref}
\modulolinenumbers[5]
\usepackage{amssymb} 
\usepackage{amsmath} 
\usepackage{stmaryrd} 
\usepackage{geometry} 
\usepackage{graphicx}
\usepackage[T1]{fontenc}
\usepackage{subfig}
\usepackage{todonotes}
\usepackage{sectsty}
\sectionfont{\fontsize{12}{8}\selectfont}
\subsectionfont{\fontsize{12}{8}\selectfont}
\subsubsectionfont{\fontsize{12}{8}\selectfont}

\journal{International Journal for Numerical Methods in Engineering}

\begin{document}

\begin{frontmatter}
\title{Local uncovering of unresolved physics in structural mechanics: seamless choice of modelling resolution using a CutFEM level-set approach}

\author[1]{E. Mikaeili\corref{aaa}}
\ead{mikaeilie@cardiff.ac.uk}
\author[1,2]{P. Kerfriden\corref{aaa}}
\ead{pierre.kerfriden@mines-paristech.fr}
\author[3]{S. Claus\corref{aaa}}
\ead{susanne.claus@onera.fr}

\cortext[aaa]{Corresponding Authors.}

\address[1]{Cardiff University, School of Engineering, \\ The Parade, CF243AA Cardiff, United Kingdom }

\address[2]{MINES ParisTech, PSL Research University, MAT--Centre des Matériaux, \\ CNRS UMR 7633, BP 87 91003 Evry, France}

\address[3]{ONERA, Université Paris-Saclay, 8 Chemin de la Huni\`{e}re, 91120 Palaiseau, France }

\begin{abstract}

In this paper, we present a robust and efficient unfitted concurrent multiscale method for continuum-continuum coupling, based on the Cut Finite Element Method (CutFEM). The computational domain is defined using approximate signed distance functions over a fixed background mesh and is decomposed into microscale and macroscale regions using a novel zooming technique. The zoom interface is represented by a signed distance function intersecting the computational mesh arbitrarily. The mesh inside the zoomed region is hierarchically refined to resolve the microstructure. In the examples considered, the microstructure may contain voids and hard inclusions, and its geometry is defined implicitly by a signed distance function interpolated over the refined mesh.

Our zooming technique allows the zoom interface to intersect the microstructure interface in an arbitrary fashion, enabling greater flexibility and accuracy in the modelling of complex geometries. The micro and macro regions are coupled using Nitsche's method, ensuring stability and accuracy in the solution. Ghost penalty terms are utilised to ensure the stability of cut elements along the zoom interface and the microstructure interface.

To demonstrate the effectiveness of our framework, we apply it for modelling several heterogeneous structures with both linear elasticity and plasticity constitutive behaviours. The results show that our framework is robust and efficient, producing accurate and reliable solutions for such problems. Our proposed method provides a highly versatile and effective approach to multiscale modeling of structures with complex microstructures, and has the potential to be extended to problems requiring seamless moving of zooming region(s) during the simulation, such as damage growth and fracture propagation.


\end{abstract}

\end{frontmatter}


\section{Introduction}

Numerical analysis of heterogeneous materials such as composites is conventionally carried out with properties obtained from homogenisation methods (see, e.g. \cite{tadmor_miller_2011, Kanoute.09, FEYEL03}), passing data from small (micro) to large (macro) length scales, wherein the macroscale properties are obtained by averaging stresses and strains over a representative volume element (RVE). However, these homogenisation methods suffer from drawbacks, including macroscopic uniformity and RVE periodicity assumptions. The uniformity assumption is not satisfied in critical regions of high gradients, such as interfaces, complex geometries with sharp angles, and regions experiencing severe plasticity and softening. The periodicity assumption is also not fulfilled when the material's microstructure is nonuniform. In this context, using direct numerical simulations leads to accurate responses, but it is not tractable for large-scale structures. Over the last few decades, such issues have been tackled effectively by domain decomposition methods (DDM) \cite{tallec1994domain, toselli2004domain}, which divide the computational domain into subdomains. One subdomain typically contains a critical phenomenon undergoing complex local mechanisms, such as damage, while the surrounding subdomain is usually homogeneous and elastic. Submodelling approach \cite{zohdi1996hierarchical, ZOHDI19992507, ZOHDI2001} is a well-known class of DDM that utilises a regularised approximation based on macroscale solutions as a boundary condition for the microscale model. Herein, the communication between two decoupled subdomains is one-way, occurring from macroscale to microscale.

Later on, the submodelling technique was advanced with concurrent solvers, where the subdomains were computed simultaneously. This category of DDM is called the "concurrent multiscale method" \cite{RAGHAVAN2004497, Dhia.Rateau.05}, suggesting coupling operators for linking the subdomains, whereby each subdomain possesses a different length-scale. In the concurrent multiscale methods, the constraints related to the coupling operators are typically applied over the interfaces between subdomains strongly, e.g. within Lagrange multiplier technique \cite{BURMAN20102680, Ji04, dsouza21} or weakly by Nitche's method \cite{BURMAN2012328, CAI2021113880,liu2014non}.

Concurrent multiscale methods face two significant challenges, as noted by \cite{Akbari15}: (i) adequately modelling the coupling between scales, and (ii) discretising the corresponding computational domain, including the subdomains and their interfaces. To address the first challenge, there are two categories of solutions available. The first category is based on overlapping techniques (also called "handshake" approaches), which handle the difficulty of coupling two models with incompatible kinematics over a region where the two models/scales are overlapped. Examples of well-known overlapping techniques include the bridging domain method \cite{Xiao2004, Tadmor1996, Beex.Kerfriden.ea.14} and the Arlequin method \cite{Dhia.98}. The second category is based on non-overlapping techniques that couple the two models over an interface, such as the mortar method \cite{Lamichhane.Wohlmuth.04}. Both overlapping and non-overlapping techniques can link subdomains with the same or different scales/physics.

In addressing the second challenge, there are two main approaches available: fitted and unfitted techniques. Fitted discretisation techniques require the edges of the elements to align with the interfaces, which can be a time-consuming and error-prone process. This often creates meshing obstacles, particularly when dealing with problems with non-stationary or time-dependent spatial properties such as microstructure and zooming geometries, which may require repeated remeshing. In contrast, unfitted techniques handle geometric descriptions independently for concurrent multiscale simulations. Well-known examples of unfitted methods include the cut finite element method (CutFEM) \cite{Burman.Claus.ea.15}, the finite cell method (FCM) \cite{Parvizian07, duster2008finite}, and the Cartesian grid finite element method (cgFEM) \cite{nadal2013efficient}. These techniques are particularly appealing for modelling problems with complex and/or time-dependent geometries. Another group of unfitted methods, extensively developed for domain decomposition analysis, includes the mortar method \cite{bernardi1989new, arbogast2000mixed}, the dual mortar method \cite{wohlmuth2000mortar}, and the localised mortar method \cite{park2002simple}. These unfitted methods generally employ Lagrange multipliers to impose weak continuity conditions over the domain decomposition interface. For their application in concurrent multiscale analysis, see studies by Arbogast et al. \cite{arbogast2007multiscale} and Subber et al. \cite{subber2016asynchronous}. In this paper, we focus on the CutFEM technique to develop an unfitted discretisation framework that deals simultaneously with complex microstructures and zooming interfaces.


The CutFEM technique can be considered as a variation of the eXtended Finite Element Method (XFEM) \cite{Moes99}, as a popular unfitted FEM for embedded interfaces. While XFEM enriches the nodes that are intersected by the interface with additional degrees of freedom using enrichment functions, CutFEM employs an overlapping fictitious domain technique to add extra elements overlapping the intersected elements. This approach allows each overlapping element to represent a different side of the interface, effectively doubling the degrees of freedom in the intersected elements. Thus, CutFEM can capture any type of discontinuity in the solution field, including strong and weak discontinuities, seamlessly. The two overlapped elements are then glued together using Nitsche's method. To ensure the stability of the intersected elements within the interface, the ghost penalty regularisation technique is often used, particularly when a small cut has been created \cite{Burman.10, hansbo2002unfitted}. The CutFEM has the advantage of being able to easily handle arbitrary and complex geometries, whereas XFEM may require additional work to handle such geometries. Another advantage of CutFEM over XFEM is that it does not require a partition of unity, which can be computationally expensive in 3D simulations.

The literature on CutFEM, similar to the XFEM (cf.\cite{becker2009nitsche, talebi2013molecular, Mikaeili18, bordas2023partition}), is vast. It has been applied for problems in two-phase fluid flow \cite{Claus19.2, FRACHON201977}, multi-physics \cite{hansbo2016cut, Claus.Bigot.ea.18}, contact mechanics \cite{Claus.Kerfriden.18, Claus21Digital} fracture mechanics \cite{poluektov2022cut}, etc. In the area of domain decomposition and multiscale modelling, one of the most relevant works is carried out by authors in \cite{JOHANSSON2019672, DOKKEN2020113129, JOHANSSON2020b}, who developed a multi-mesh framework based on CutFEM for multi-component structures. In their methodology, each component of the large structure is meshed separately while each mesh is allowed to overlap the CutFEM fixed background mesh arbitrarily. Their method allows the CutFEM background mesh to be intersected by more than one interface (corresponding to the overlapped meshes) simultaneously. Moreover, they utilise Nitche's method to enforce interface conditions in the intersected elements. In the context of concurrent multiscale modelling, however, to the author's best knowledge, there are no published papers by other authors using the CutFEM technique.

In our prior publication \cite{mikaeili2022concurrent}, we developed a concurrent micro-macro model blending method which uses one background mesh to describe the multiscale problem. The macroscale region has been represented by a homogenised material over coarse elements, and the microscale structure is represented by a signed distance function and has been discretised using a CutFEM approach over very fine elements. The macro and micro regions have been coupled via a mixing region with a thickness of several mesh elements.  

The multiscale framework, which is presented in this article, is different from our previous mixing approach in two significant ways. Firstly, the multiscale approach presented in the following does not feature a mixing region but couples the microscale domain to the macroscale domain via a sharp interface. Secondly, we no longer consider that we have one adaptive background mesh but rather two distinct meshes and discretisations: one for the macroscale domain and one for the microscale domain. The macroscale domain is modelled as a homogenised domain and represented by a coarse mesh. The microstructure is discretised independently from the macro domain and features rich geometrical detail. The microscale domain is assumed to exist everywhere in the macroscale domain. Zoom regions are then defined in areas of interest on the macro scale domain to open "windows" into the microscale region. These zoom regions are dynamic and can change over time. These dynamic zoom regions are beneficial for problems which feature sharply localised phenomena, such as crack propagation or the formation of plastic bands. In this article, we will demonstrate the capabilities of our method for the latter (non-linear plasticity model).  

Within our framework, we first define regions of interest or "zooms" implicitly through a level set function interpolated over a fixed coarse background mesh (discretised macroscale domain). The corresponding zoom interface can intersect the background mesh arbitrarily. A second level set function is introduced to define the microstructure, which is interpolated over a high-resolution mesh. The macro and micro regions are glued together using Nitsche's method. Then to guarantee the well-conditioning of the multiresolution system matrix and the stability of the solver, cut elements are regularised with the ghost penalty technique.

The paper is outlined as follows. In section \ref{Section2}, we will present strong and weak forms of governing equations for the microscale and macroscale models within concurrent multiscale formulation. Then, we will discretise the corresponding formulation using the CutFEM technique. In section \ref{Section3}, the proposed multiscale framework will be firstly validated and then tested for different heterogeneous structures with linear elasticity and plasticity behaviours.

\section{Concurrent multiscale model using zooms described by level-set functions}
\label{Section2}

\begin{figure}[h]
   \centering
     \subfloat[ \label{1}]{%
     \includegraphics[scale=0.29]{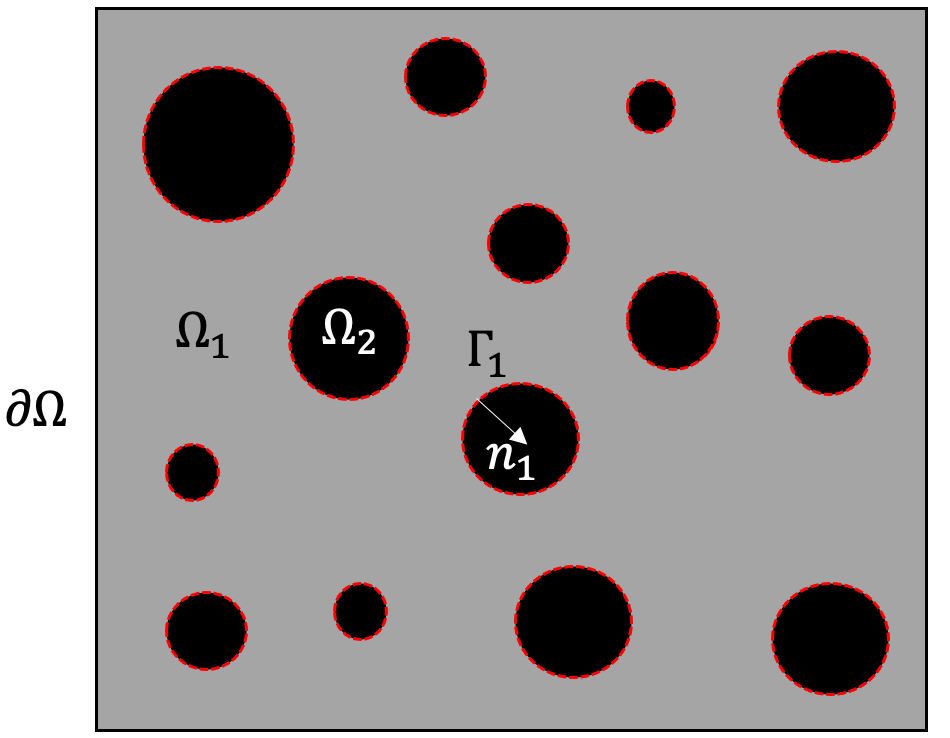}
     }
    \hfill
     \subfloat[ \label{2}]{%
      \includegraphics[scale=0.385]{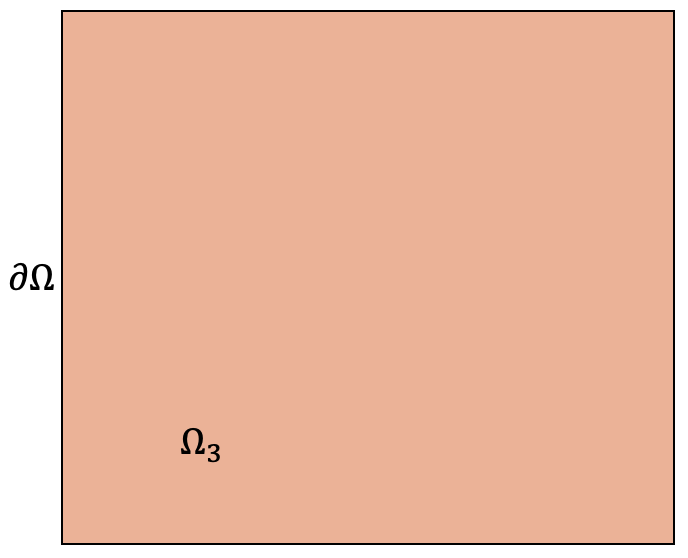}
     }
    \hfill
        \subfloat[ \label{3}]{%
     \includegraphics[scale=0.385]{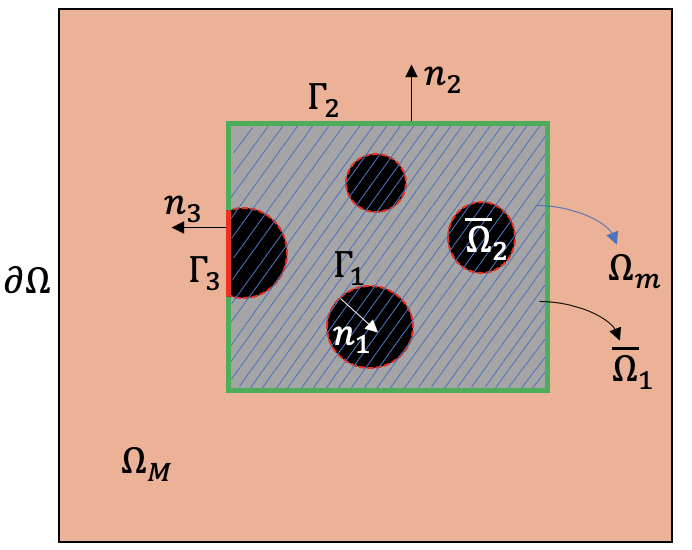}
     }
    \hfill
     \caption{Domain partitioning: (a) heterogeneous microscale model, (b) homogeneous macroscale model and (c) multiscale model (the macroscale model enriched with a zoom including microscale model)}
     \label{fig:schmm}
\end{figure}


In this section, we commence by defining the spatial domains relevant to our concurrent multiscale framework in Section \ref{Formulation:Domains}. The schematic representations of these domains are depicted in Figure \ref{fig:schmm}(a) and (b). Subsequently, we present the strong form of the governing equations for the heterogeneous microscale model in Section \ref{Formulation:StrongFHeter}, followed by the equivalent homogenised macroscale model in Section \ref{Formulation:StrongFHomog}. Next, we enrich the homogenised macroscale model with microscale features within regions of interest (zooms), which forms the basis for our multiscale framework. The domain partitioning for the multiscale model is then illustrated in Figure \ref{fig:schmm}(c). We derive the weak form of the governing equations for the proposed concurrent multiscale model in Section \ref{Formulation:WeakF}, while employing Nitsche's formulation to couple the subdomains across sharp interfaces. We proceed to define the corresponding subdomains and interfaces in terms of level set functions in Section \ref{Formulation:DomainsLevelsets}. Finally, in Section \ref{Formulation:Discretisation}, we discretise the macro and microscale problems using CutFEM.


\subsection{Multiscale domain with zoom regions}
\label{Formulation:Domains}

Let us consider that we have a two-phase composite material occupying domain $\Omega$ with boundary $\partial \Omega$, which consists of a matrix-phase $\Omega_1$ and an inclusion phase $\Omega_2$ (see Figure~\ref{fig:schmm} (a)). Here, $\Omega_1 \cup \Omega_2 = \Omega$ and $\Omega_1 \cap \Omega_2 = \emptyset$. The interface between $\Omega_1$ and $\Omega_2$ is denoted by $\Gamma_1$. Furthermore, let us assume that the two-phase composite can be represented by a homogenised material, which we denote by $\Omega_3$ (see Figure~\ref{fig:schmm} (b)). \\ 
Now, let us introduce zooms in the homogenised domain into the two-phase composite material. The zoom regions are denoted by $\Omega_m = \bar{\Omega}_1 \cup \bar{\Omega}_2$, where $\bar{\Omega}_i$, $i=1,2$, denotes the part of $\Omega_i$ that lies inside the zoom regions (microscale domain). And ${\Omega}_\mathcal{M} = \Omega_3 \setminus \Omega_m$ denotes the homogenised region without the zoom regions (macroscale domain). The interface between the matrix phase of the zoom regions and the macroscale domain is denoted by $\Gamma_2= \bar{\Omega}_1 \cap \bar{\Omega}_3$ and the interface between the inclusion phase and the macroscale domain is denoted by $\Gamma_3 = \bar{\Omega}_2 \cap \bar{\Omega}_3$ (see Figure~\ref{fig:schmm} (c)). \\
We then consider

\subsection{Heterogeneous elasticity problem: strong form}
\label{Formulation:StrongFHeter}

\subsubsection{Semi-discrete boundary value problem}

 The semi-discrete problem of elasticity that we wish to solve is the following. Time interval $\mathcal{I} = [0 \, T]$ into $N$ equally spaced time intervals. At discrete times in $\mathcal{I}_{\Delta T} = \{ t_1=\Delta T , \, t_2=2 \Delta T , \, ... \, ,  t_N= N \Delta T = T \}$, we look for displacement $ u ^n :=  \{ u^{n} _1, u^{n} _2\} : \Omega_1 \times \Omega_2 \rightarrow \mathbb{R}^D \times \mathbb{R}^D$ at $n^{\text{th}}$ time step satisfying, 
\begin{equation}
\forall \, i \in \{ 1,2\}, \qquad \text{div} \, \sigma^{n} _i( \nabla_s u_i ^n) + f^n = 0 
\qquad \text{in} \, \Omega_i \, 
\end{equation}
where $\sigma^n_i(\nabla_s u_i ^n)$ is the stress tensor, which is a function of strain tensor $\nabla_s u_i ^n$, and $f^n$ is the corresponding body force. 

The boundary conditions of the elasticity problem are
\begin{equation}
\label{eq:DBC}
\forall \, i \in \{ 1,2\}, \qquad u^{n} _i = u^n_d \qquad \text{over} \ \partial \Omega_u \cap\partial \Omega_i 
\end{equation}
and
\begin{equation}
 \forall \, i \in \{ 1,2\}, \qquad \sigma_i ^n(\nabla_s u^{n} _i) \cdot n_{\partial \Omega} = \tau^n  
 \qquad \text{over} \ \partial \Omega_t \cap \partial \Omega_i 
\end{equation}
where $\tau ^n$ denotes the traction vector and $n_{\partial \Omega}$ is the outer normal to the boundary. Here, $\partial \Omega = \partial \Omega_t  \cup \partial \Omega_u$, $ \partial \Omega_t  \cap \partial \Omega_u = \{ 0 \}$. Moreover, fields $f^n$, $u^n_d$ and $\tau ^n$ are given time-discrete fields. \\


\subsubsection{Linear elasticity}

If we assume that the two phases of the composite are linear elastic, time-independent and homogeneous, the stress functions $\sigma^{n} _i$ may be expressed as
\begin{equation}
\forall \, i \in \{ 1,2\}, \qquad \sigma^{n} _i(\nabla_s u^{n} _i) := C_i :\nabla_s u^{n} _i \qquad \text{in} 
\, \Omega_i  \, ,
\end{equation}
where $\nabla_s \, . \,  =\frac{1}{2} (\nabla \, . \, +\nabla^T \, . \, ) $ and $C_i$ is the fourth-order Hooke tensor of the material occupying phase $i$. This tensor may be expressed as a function of the the Lam\'e coefficients $\lambda_i$ and $\mu_i$ as follows:
\begin{equation}
\forall \, i \in \{ 1,2\}, \qquad C_i : \nabla_s \, . = \lambda_i \text{Tr} ( \nabla_s \, .  \, )  \mathbb{I} + 2 \mu_i \nabla_s \, . \, .
\end{equation}





\subsubsection{von Mises plasticity}

\paragraph{Time continuous constitutive law.}

We consider the following von Mises plasticity model. The stress $s$ in the material is given by
\begin{equation}
    s = C: ( \epsilon - \epsilon_p)
\end{equation}
as a function of the strain $\epsilon$ (\textit{i.e.} the symmetric part of the displacement gradient) and the plastic strain $\epsilon_p$. The yield surface is defined as
\begin{equation}
    f(s_D,q,p)  = \sqrt{ \frac{3}{2} (s_D-q) : (s_D-q)  } - (Y_0 + R(p)) 
\end{equation}
where $s_D$ denotes the deviatoric part of $\sigma$, $p$ denotes the cumulative plastic strain and $q$ the back stress. In equation above, $Y_0$ is a constant initial yield parameter, the term $R(p) = \hat{H} \, p$ is the isotropic linear hardening function, where $\hat{H}$ is the corresponding hardening modulus. Moreover, the plasticity flow rules are as follows:
\begin{equation}
  \lambda \geq 0 \, , \qquad \lambda f(s_D,q,p)=0 \, , \qquad f(s_D,q,p)  \leq 0
\end{equation}
where
\begin{equation}
\begin{array}{l} \displaystyle 
    \dot{\epsilon}_p  = \lambda \left( \frac{s_D-q}{ \| s_D-q \|} \right) \, ,
    \\  \displaystyle 
     \dot{q} = \lambda  \left( \bar{H} : \frac{s_D-q}{ \| s_D-q \|} \right) \, ,
     \\  \displaystyle 
      \dot{p} = \lambda \, .
\end{array}
\end{equation}

In equations above, $\lambda$ is the plastic multiplier and $\bar{H}$ is a fourth-order kinematic hardening tensor. In our examples, $\bar{H}$ will be vanishingly small (no kinematic hardening).

\paragraph{Implicit time integration}

The previous ODE may be discretised in time using an implicit Euler scheme. This leads to the following semi-discrete material law.
\begin{equation}
\begin{array}{l} \displaystyle 
\frac{1}{\Delta T} \left( C^{-1}:s^{n+1} - \epsilon^{n+1} + \epsilon^n_p \right) 
+ \lambda  \, \frac{s_D^{n+1}-q^{n+1}}{ \| s_D^{n+1}-q^{n+1} \|}  = 0
   \\  \displaystyle 
\frac{1}{\Delta T} \left( \bar{H}^{-1}: (q^{n+1} - q^n)  \right) + \lambda  \, \frac{s_D^{n+1}-q^{n+1}}{ \| s_D^{n+1}-q^{n+1} \|}  = 0
    \\  \displaystyle 
\lambda \geq 0\, , \qquad  \lambda f(s_D^{n+1},q^{n+1},p^{n}+ \lambda \Delta T) = 0\, , \qquad f(s_D^{n+1},q^{n+1},p^{n}+ \lambda \Delta T )  \leq 0 \, .
\end{array}
\end{equation}

Given $( \epsilon^{n+1} , \epsilon_p^n, q^{n},p^n ) $,
the previous nonlinear system of equation (the last three constraints can be recast as a single nonlinear equality using the Heaviside function) can be solved for $(s^{n+1},q^{n+1},\lambda)$ 
using the usual combination of operator splitting and Newton iterative solution scheme. The update of internal variables is performed according to
\begin{equation}
\begin{array}{l} \displaystyle 
\epsilon^{n+1}_p  = \epsilon^{n}_p + \lambda \left( \frac{s^{n+1}_D-q^{n+1}}{ \| s^{n+1}_D-q^{n+1} \|} \right) \Delta T 
\\  \displaystyle
p^{n+1}= p^{n} + \lambda \Delta T 
\end{array}
\end{equation}

The procedure can therefore be summarised as an (implicit defined) relationship 
\begin{equation}
    s^{n+1} =   s_{\Delta T} \left( \epsilon^{n+1} , (\epsilon_p^n, q^{n},p^n ) ; \mu , \Delta T \right)
\end{equation}
where $\mu$ is a real-valued vector containing all the parameters of the constitutive law: $Y_0$, all the free parameters of tensor $\bar{H}$ and that of $R$.

\paragraph{Semi-discrete implicit stress functions}

At time $t_n > 0$, for any phase index $i \in \llbracket 1 \, 2 \rrbracket$ of the composite material, we may replace the elastic constitutive law by the following nolinear function
\begin{equation}
\sigma^{n} _i (\nabla_s u^{n} _i) = 
 s_{\Delta T} \left( \nabla_s u^{n} _i , \xi^{n-1} _i ;\mu_i , \Delta T \right)
\end{equation}
where the field of past internal variables $\xi^{n-1} _i = ( \epsilon_p^{n-1}, q^{n-1}, p^{n-1} )$ defined over $\Omega_i$ are sequentially and locally  updated according to the procedure outlined above. We suppose that at the beginning of the simulation, all the internal variables are null. \\


\subsection{Homogenised elasticity problem: strong form}
\label{Formulation:StrongFHomog}

Here, we present a homogenised elasticity problem with an equivalent homogenised material for the two-phase composite material defined in the previous section. The corresponding computational domain is shown schematically in Figure \ref{fig:schmm} (b). Similar to the heterogeneous elasticity problem, we consider a time-dependent problem with time interval $\mathcal{I}=[0 \ T]$ divided into $N$ equal time steps. In each time step $n$, we look for displacement $u^n _3 : {\Omega}_3 \rightarrow \mathbb{R}^D$ satisfying the following equilibrium equation

\begin{equation}
\text{div} \, \sigma^{n}_3( \nabla_s u^n _3) + f^n_3 = 0 
\qquad \text{in} \, \, \, \, \Omega_3 \, ,
\end{equation}
where the corresponding boundary conditions are as follows: $u_3 ^n = u^n _d$ over $\partial\Omega_u \cap \partial\Omega$ and $\sigma_3 ^n .n_{\partial\Omega}=t_d ^n$ over $\partial\Omega_n \cap \partial\Omega$, where $\partial\Omega = \partial\Omega_t \cup \partial\Omega_u$ and $\partial\Omega_t \cap \partial\Omega_u = \{0\}$.

In $\Omega_3$, we introduce a surrogate material model with slowly varying parameters in space. If the coarse material is elastic and homogeneous, it is characterised by constant tensor $C _3$ (which may be obtained, for instance, via some form of homogenisation of the composite material), whose action reads as

\begin{equation}
\qquad C_3 : \nabla_s \, . = \lambda_3 \text{Tr} ( \nabla_s \, .  \, )  \mathbb{I} + 2 \mu^c \nabla_s \, . \,
\end{equation}
The resulting stress function is

\begin{equation}
 \sigma_3(\nabla_s u_3) := C_3 :\nabla_s u_3 \qquad \text{in}  \ \Omega_3 \, .
\end{equation}

If the coarse material is plastic, we define the associated stress update at the $n^{th}$ time increment, $n \in \{ 1,2, ..., N \}$, by

\begin{equation}
\sigma^{n} _3 (\nabla_s u^{n} _3) = 
  s_{\Delta T} \left( \nabla_s u^{n} _3 , \xi^{n-1} _3 ; \mu_3 , \Delta T \right).
\end{equation}

The solution approach for equation above is similar to the approach presented for heterogeneous plasticity problems in the previous section.



\subsection{Multiscale problem in weak form}
\label{Formulation:WeakF}

Keeping this in mind, we now split domain $\Omega$ arbitrarily into two non-overlapping domains: the coarse domain $\Omega_\mathcal{M} =: \Omega_3$ and the fine domain $\Omega_m$, as shown in Figure \ref{fig:schmm} (c). 
Let us now redefine 
$\Omega_{i}$ as $\Omega_m \cap \Omega_i$, for $i \in \{ 1,2\}$ (level set $\phi_1$ is unaffected by this change of notation).
The interfaces between the domains 
\begin{equation}
\begin{aligned}
\Gamma_1 = \Omega_1 \cap \Omega_2\, , \\
\Gamma_2 = \Omega_1 \cap \Omega_3\, , \\
\Gamma_3 = \Omega_2 \cap \Omega_3.
\end{aligned}
\end{equation}

In the following, we present the weak form of governing equations for the microscale and macroscale problems separately. However, due to the concurrent multiscale modelling, both problems will be solved simultaneously. 

\subsubsection{Microscale model in weak form}

Now, the microscale problem of elasticity at time $t_n \in \mathcal{I}_{\Delta T}$ reads: We look for a displacement field $u_m:= \{u_1,u_2\}: \bar{\Omega}_1 \times \bar{\Omega}_2 \rightarrow \mathbb{R}^D \times \mathbb{R}^D$  satisfying

\begin{equation}
    a_m(u_m, \delta u_m) + a_{m,\sharp} (u_m, \delta u_m) = l_m(\delta u_m) \, .
    \label{EQ:multires.micro.weakf}
\end{equation}

In the previous variational statement, arbitrary $\delta u_m:= \{u_1,u_2\}: \bar{\Omega}_1 \times \bar{\Omega}_2 \rightarrow \mathbb{R}^D \times \mathbb{R}^D$  is required to satisfy the homogeneous Dirichlet conditions
\begin{equation}
 \delta u_m = 0 \qquad \text{over} \qquad \partial \Omega_{m,u} := \partial \Omega_u \cap\partial \Omega_m \, .
\end{equation}

In equation \ref{EQ:multires.micro.weakf}, the bilinear form $a_m$ is defined as,
\begin{equation}
 a_m(u_m, \delta u_m) = \int_{\Omega_m} \sigma_m( \nabla_s u_m) : \nabla_s \delta u_m \, dx \, . 
\end{equation}
The corresponding linear form $l_m$ is given by,
\begin{equation}
 l_m(\delta u_m) = \int_{\Omega_m} f \cdot \delta u_m \, dx +   \int_{\partial \Omega_{m,t}} \tau \cdot \delta u_m  \, dx \,  , 
\end{equation}
where
\begin{equation}
 \Omega_{m,t} = \Omega_m \cap \partial \Omega_t \, .
\end{equation}

The bilinear form $a_{m,\sharp}$ in equation \ref{EQ:multires.micro.weakf} introduces the coupling terms related to the Nitsche's method to glue together the microscale subdomains, i.e. $\bar{\Omega}_1$ and $\bar{\Omega}_2$, and is expressed as,

 \begin{equation}
 \begin{array}{rcl}
\displaystyle a_{m,\sharp}(u_m,\delta u_m) 
& = & \displaystyle \gamma_1 \hat{w}_1
\int_{\Gamma_1}  \llbracket u_m \rrbracket_1 \cdot \llbracket \delta u_m \rrbracket_1 \, dx
\\
& - & \displaystyle 
\int_{\Gamma_1}  \left\{ t \right\}_1(u_m) \cdot  
\llbracket \delta u_m \rrbracket_1  \, dx
\\
& - & \displaystyle 
\int_{\Gamma_1}  \left\{ t \right\}_1( \delta u_m) \cdot  
\llbracket u_m \rrbracket_1  \, dx \, .
 \end{array}
 \label{EQ:multires.Nitsche.microW}
\end{equation}
where 
\begin{equation}
 \llbracket u_m \rrbracket_1 = u_1 - u_2 \, ,
\end{equation}
which denotes the jump in the displacement field across $\Gamma_1$.
In equation \ref{EQ:multires.Nitsche.microW}, $\gamma_1 > 0$, $\hat{w}_1$ and $\left\{ t \right\}_1$ are Nitsche's parameters for the microscale model.

\subsubsection{Macroscale model in weak form}

Now, for the macroscale problem of elasticity at time $t_n \in \mathcal{I}_{\Delta T}$ reads, we look for a displacement field $u_{\mathcal{M}}: {\Omega}_{\mathcal{M}}  \rightarrow \mathbb{R}^D $ satisfying

\begin{equation}
a_{\mathcal{M}}(u_{\mathcal{M}}, \delta u_{\mathcal{M}}) + a_{{\mathcal{M}}, \sharp} (u_{\mathcal{M}}, \delta u_{\mathcal{M}})  = l_{\mathcal{M}}(\delta u_{\mathcal{M}})
\label{Eq:MacroWeakF}
\end{equation}
where the arbitrary triplet $\delta u_{\mathcal{M}}:  \Omega_{\mathcal{M}} \rightarrow \mathbb{R}^D$ is required to satisfy the homogeneous Dirichlet conditions
\begin{equation}
 \delta u_{\mathcal{M}} = 0 \qquad \text{over} \qquad \partial \Omega_{{\mathcal{M}},u} := \partial \Omega_u \cap\partial \Omega_{\mathcal{M}} \, . 
\end{equation}

The bilinear form $a_{\mathcal{M}}$ in equation \ref{Eq:MacroWeakF} is defined as follows,
\begin{equation}
a_{\mathcal{M}}(u_{\mathcal{M}}, \delta u_{\mathcal{M}}) = \int_{\Omega_{\mathcal{M}}} \sigma_{\mathcal{M}}( \nabla_s u_{\mathcal{M}}) : \nabla_s \delta u_{\mathcal{M}} \, dx, 
\end{equation}
The corresponding linear form $l_{\mathcal{M}}(u_{\mathcal{M}}, \delta u_{\mathcal{M}})$ is defined as,
\begin{equation}
l_{\mathcal{M}}(\delta u_{\mathcal{M}}) = \int_{\Omega_{\mathcal{M}}} f \cdot \delta u_{\mathcal{M}} \, dx + \int_{\partial \Omega_{{\mathcal{M}},t}} \tau \cdot \delta u_{\mathcal{M}}  \, dx \, , 
\end{equation}
where
\begin{equation}
\partial \Omega_{{\mathcal{M}},t} = \Omega_{\mathcal{M}} \cap \partial \Omega_t \, .
\end{equation}

The bilinear form $a_{\mathcal{M}, \sharp} (u_{\mathcal{M}}, \delta u_{\mathcal{M}})$ presents the Nitsche's formulation for coupling microscale and macroscale models over two types of interfaces; $\Gamma_2$ and $\Gamma_3$, which is denoted by
\begin{equation}
a_{\mathcal{M}, \sharp} (u_{\mathcal{M}}, \delta u_{\mathcal{M}}) = a_{2,\sharp} (u_{\mathcal{M}}, \delta u_{\mathcal{M}}) + a_{3,\sharp} (u_M, \delta u_{\mathcal{M}})
\end{equation}
where for $i \in \{ 2, 3 \}$, the Nitsche's formulation over $\Gamma_i$ reads
\begin{equation}
 \begin{array}{rcl}
\displaystyle  a_{i, \sharp} (u_{\mathcal{M}}, \delta u_{\mathcal{M}}) 
& = & \displaystyle \gamma_i \hat{w}_i
\int_{\Gamma_i}  \llbracket u_{\mathcal{M}} \rrbracket_i \cdot \llbracket \delta u_{\mathcal{M}} \rrbracket_i \, dx
\\
& - & \displaystyle 
\int_{\Gamma_i}  \left\{ t \right\}_{i}(u_{\mathcal{M}}) \cdot  
\llbracket \delta u_{\mathcal{M}} \rrbracket_i  \, dx
\\
& - & \displaystyle 
\int_{\Gamma_i}  \left\{ t \right\}_{i}( \delta u_{\mathcal{M}}) \cdot  
\llbracket u_{\mathcal{M}} \rrbracket_i  \, dx \, .
 \end{array}
\end{equation}
where $\gamma_i > 0$, $\hat{w}_i$ and $\{ t \}_{i}$ are Nitsche's terms and we have
\begin{equation}
 \llbracket u_{\mathcal{M}} \rrbracket_i = u_{\mathcal{M}} - u_i \, ,
\end{equation}
which denotes the jump in the displacement field across $\Gamma_i$.

\subsection{Level-set-based descriptions of subdomains and interfaces}
\label{Formulation:DomainsLevelsets}

In this contribution, we define two subdomains, $\Omega_1$ and $\Omega_2$ implicitly using a time-independent continuous level set function $\phi_1 \in \mathcal{C}^0(\Omega)$. The subdomains are given as follows:

\begin{equation}\begin{array}{rcl}
\Omega_1 = \{ x \in \Omega \, | \, \phi_1 (x) \leq 0  \}
\\
\Omega_2 = \{ x \in \Omega \, | \, \phi_1 (x) > 0  \}
\end{array} \, .
\end{equation}

Furthermore, the interface separating these two subdomains is denoted by $\Gamma_1$ and defined in terms of $\phi_1$ as follows:

\begin{equation}
 \Gamma_1 =   \{ x \in \Omega \, | \, \phi_1 (x) = 0  \} \, .
\end{equation}

The coarse domain, denoted by $\Omega_\mathcal{M}$, is defined using another continuous level set function $\phi_2 \in \mathcal{C}^0(\Omega)$, as:

\begin{equation}
    \Omega_\mathcal{M} = \{ x \in \Omega \, | \, \phi_2 (x) > 0  \} \, .
\end{equation}

Similarly, the fine domain, denoted by $\Omega_m$, is defined as:

\begin{equation}
    \Omega_m = \{ x \in \Omega \, | \, \phi_2 (x) \leq 0  \} \, ,
\end{equation}
where $\Omega_m$ is comprised of micro pores (or micro inclusions) and matrix. Specifically, the subdomain corresponding to the micro pores or inclusions inside $\Omega_m$ is defined as:

\begin{equation}
    \bar{\Omega}_1 = \{ x \in \Omega \, | \, \phi_1 (x) \geq 0, \ \ \phi_2 (x)\leq 0 \} \, .
\end{equation}

On the other hand, the subdomain corresponding to the matrix inside $\Omega_m$ is defined as:

\begin{equation}
    \bar{\Omega}_2 = \{ x \in \Omega \, | \, \phi_1 (x) \leq 0 , \ \ \phi_2 (x)\leq 0 \  \} \, .
\end{equation}

Finally, we introduce the interface between different subdomains defined above. The interface between the coarse and fine domains, denoted by $\Gamma_{12}$, is expressed as:

\begin{equation}
 \Gamma_{12} =   \{ x \in \Omega \, | \, \phi_2 (x) = 0  \} \, ,
\end{equation}
which is comprised of two separate interfaces: $\Gamma_{2}$ and $\Gamma_{3}$, defined as:

\begin{equation}\begin{array}{rcl}
\Gamma_2 = \{ x \in \Gamma_{12} \, | \, \phi_1 (x) > 0  \}
\\
\Gamma_3 = \{ x \in \Gamma_{12} \, | \, \phi_1 (x) \leq 0  \}
\end{array} \, .
\end{equation}

\subsection{Discretisation of the multiresolution problem}
\label{Formulation:Discretisation}

\subsubsection{Discretisation of the geometry}

Let us introduce a coarse triangulation $\mathcal{T}^H$ of domain 
$\Omega$. The tessellated domain is denoted by $\Omega^H$. Furthermore, let us introduce the finite element space of continuous piecewise linear functions, \textit{i.e.}

\begin{equation}
\mathcal{Q}^H : = \{ w \in  \mathcal{C}^0(\Omega^H): w |_{K} \in \mathcal{P}^1(K) \, , \forall K \in \mathcal{T}^H \} \, .
\end{equation}

We now define the finite element approximation of  coarse domain $\Omega_\mathcal{M}$ as

\begin{equation}
    \Omega^H _\mathcal{M} =\Omega^H _3 = \{ x \in \Omega^H \, | \, \phi_2 ^H(x) \geq 0  \} \, ,
\end{equation}
where $\phi_2 ^H(x) \in \mathcal{Q}^H$ is the coarse nodal interpolant of $\phi_2$. 

Let us now introduce a hierarchical subtriangulation $\mathcal{T}^h$ of $\mathcal{T}^H$, with $h \ll H$. Due to the hierarchical structure, the union of all triangles of $\mathcal{T}^h$ is the coarse finite element domain $\mathcal{T}^H$.
We define space
\begin{equation}
\mathcal{Q}^{(H,h)} : = \{ w \in  \mathcal{C}^0(\Omega^H): w |_{K} \in \mathcal{P}^1(K) \, , \forall K \in \mathcal{T}^h \} \, .
\end{equation}

With this definition, domains $\Omega_1$ and $\Omega_2$ are discretised as follows.
\begin{equation}\begin{array}{rcl}
\Omega^{(H,h)} _1 = \{ x \in \Omega^H \, | \, \phi^H _2(x) \leq 0 , \, \phi^h _1(x) \leq 0 \} 
\\
\Omega^{(H,h)} _2 = \{ x \in \Omega^H \, | \, \phi^H _2(x) \leq 0 , \, \phi^h _1(x) \geq 0 \}
\end{array} \, .
\end{equation}

We define the interface between the fine domains as 
\begin{equation}
\Gamma^{(H,h)} _{1} = \{ x \in \Omega^H \, | \, \phi^H _2(x) \leq 0 , \, \phi_1 ^{(H,h)}(x) = 0  \} \, . 
\end{equation}
and the interfaces between the coarse and the fine domains as
\begin{equation}\begin{aligned}
\Gamma^H _{12} = \{ x \in \Omega^H \, | \, \phi_2 ^H(x) = 0  \} \, , \\ 
\Gamma^{(H,h)} _{2} = \{ x \in \Gamma^H _{12} \, | \, \phi_1 ^h(x) \leq 0  \} \, , \\ 
\Gamma^{(H,h)}_{3} = \{ x \in \Gamma^H _{12} \, | \, \phi_1 ^h(x) \geq 0  \} \, .\end{aligned} 
\end{equation} 

Notice that finely discretised quantities are parameterised by a pair of mesh characteristics  $ \ \mathcal{H} = (H,h)$. This is due to the hierarchical structure of the multiresolution scheme that we have introduced (the coarse domain "overshadows" the composite material). To simplify the notations, the coarse sets and variables that only depend on $H$ will also be written to be dependent on $\mathcal{H}$.

\subsubsection{Overlapping domain decomposition}

For the three different domains of the multiresolution scheme, we need to define appropriate extended domains. Such an extended domain is composed of all the elements that have a non-void intersection with its non-extended counterpart. Hence, the set of all elements of $\mathcal{T}^\mathcal{H}$ that have a non-zero intersection with $\Omega^\mathcal{H} _\mathcal{M}$ is

\begin{equation}
\hat{\mathcal{T}}^\mathcal{H} _\mathcal{M}  := \{ K \in \mathcal{T}^\mathcal{H}: K \cap \Omega^\mathcal{H} _\mathcal{M} \neq \emptyset\} \, .
\end{equation}

The fictitious domain domain corresponding to this set is $\hat{\Omega}^\mathcal{H} _\mathcal{M} := \bigcup_{K \in \hat{\mathcal{T}}^\mathcal{H} _\mathcal{M}}  K $.
Similarly for the fine domains,
\begin{equation}
\forall i \in \{ 1,2\}, \qquad \hat{\mathcal{T}}^\mathcal{H}_i  := \{ K \in \mathcal{T}^\mathcal{H}: K \cap \Omega^\mathcal{H}_i \neq \emptyset\} \, .
\end{equation}

The domains corresponding to these sets are denoted by $\hat{\Omega}^\mathcal{H}_i := \bigcup_{K \in \hat{\mathcal{T}}^\mathcal{H}_i } K $, for $i=1$ and for $i=2$.

\subsubsection{Extended interface FE spaces}

We will look for an approximation $u^\mathcal{H} = \left( u^\mathcal{H}_1, u^\mathcal{H}_2,u^\mathcal{H}_3 \right) $ of the multiresolution elasticity problem in space $\mathcal{U}^\mathcal{H} = \mathcal{U}^\mathcal{H}_1 \times \mathcal{U}^\mathcal{H}_2 \times \mathcal{U}^\mathcal{H}_3$, where
\begin{equation}
\begin{aligned}
\mathcal{U}^\mathcal{H}_3 = \mathcal{U}^\mathcal{H} _\mathcal{M}  &:= \{ w \in  \mathcal{C}^0(\hat{\Omega}^\mathcal{H}_\mathcal{M}): w |_{K} \in \mathcal{P}^1(K) \, \forall K \in \hat{\mathcal{T}}^\mathcal{H} _\mathcal{M} \} \, ,
 \\ 
\forall i \in \{ 1, 2 \} , \quad \hat{\mathcal{U}}^\mathcal{H}_i &:= \{ w \in  \mathcal{C}^0(\hat{\Omega}^\mathcal{H}_i): w |_{K} \in \mathcal{P}^1(K) \, \forall K \in \hat{\mathcal{T}}^\mathcal{H}_i \} \, .
\end{aligned}
\end{equation}
Notice that $u^\mathcal{H}$ is multi-valued in the elements that are cut by the two embedded interfaces. This feature allows us to represent discontinuities at the two interfaces.

The field of internal variables $\xi^{n}_i$, for any $n \in \llbracket 0 \, N\rrbracket$ and for any $i \in \llbracket 1 \, 3\rrbracket$, will be defined over the corresponding approximated domain ${\Omega}^\mathcal{H}_i$. These fields do not need to be extended to the fictitious domain.




\subsubsection{Additional sets}

We now define some additional sets, which is required to introduce the stabilisation strategy for our implicit boundary multiresolution formulation.

For stabilisation purpose, let us denote all elements which are intersected by $\Gamma^\mathcal{H}_{12}$ by 
\begin{equation}
\hat{\mathcal{G}}_{12}^\mathcal{H} := \{ K \in \mathcal{T}^\mathcal{H} \, | \,  K \cap \Gamma^\mathcal{H}_{12} \neq \emptyset\} \, .
\end{equation}

The domain corresponding to this set is denoted by $\hat{\Gamma}_{12}^\mathcal{H} := \bigcup_{K \in \hat{\mathcal{G}}_{12}^\mathcal{H} } K $. Similarly for the fine domains, for $i \in \{ 1, 2\}$,
\begin{equation}
\hat{\mathcal{G}}_i^\mathcal{H} := \{ K \in \mathcal{T}^\mathcal{H} \, | \,  K \cap \Gamma^\mathcal{H}_i \neq \emptyset\} \, ,
\end{equation}
and the corresponding domains will be denoted by $\hat{\Gamma}_i^\mathcal{H} := \bigcup_{K \in \hat{\mathcal{G}}_i^\mathcal{H} } K $.
We now define the set of ghost penalty element edges for 
fictitious domain $\hat{\Omega}^\mathcal{H}_1$ 
\begin{equation}
\hat{\mathcal{F}}^G_1:= \{ F = K \cap K'  : K  \in \hat{\mathcal{T}}^\mathcal{H}_1 \mbox{ and } K' \in \hat{\mathcal{T}}^\mathcal{H}_1, \, F \cap \hat{\Gamma}_1^\mathcal{H} \neq \emptyset  \} \, ,
\end{equation}
and for fictitious domain $\hat{\Omega}^\mathcal{H}_i $, $i \in \{ 2, 3\}$ as
\begin{equation}
\hat{\mathcal{F}}^G_i:= \{ F = K \cap K'  : K  \in \hat{\mathcal{T}}^\mathcal{H}_i \mbox{ and } K' \in \hat{\mathcal{T}}^\mathcal{H}_i, \, F \cap \hat{\Gamma}_i^\mathcal{H} \neq \emptyset  \} \, .
\end{equation}

\subsection{Implicit boundary finite element formulation}

The finite element multiresolution formulation is as follows: 
for any $\delta u^\mathcal{H} \in \mathcal{U}^\mathcal{H}$ satisfying the homogeneous Dirichlet boundary conditions,
\begin{equation}
    a^\mathcal{H}(u^\mathcal{H},\delta u^\mathcal{H}) + a_\sharp^\mathcal{H}(u^\mathcal{H},\delta u^\mathcal{H}) + a_\heartsuit^\mathcal{H}(u^\mathcal{H},\delta u^\mathcal{H}) = l^\mathcal{H}(\delta u^\mathcal{H}) \, \, .
\end{equation}
In the previous formulation, the bilinear form $a^\mathcal{H}$ is defined by
\begin{equation}
a^\mathcal{H}(u^\mathcal{H},\delta u^\mathcal{H}) =   \sum_{i=1}^3 \int_{\Omega^\mathcal{H}_i} \sigma_i(\nabla_s u^\mathcal{H}_i) : \nabla_s \delta u_i^\mathcal{H} \, dx \, ,
\end{equation}
and the linear form $l^\mathcal{H}$ is as follows:
 \begin{equation}
   \displaystyle l^\mathcal{H}(\delta u^\mathcal{H}) 
   =
 \sum_{i=1}^3  \int_{\Omega^\mathcal{H}_i} f \cdot \delta u^\mathcal{H}_i \, dx 
 + 
 \sum_{i=1}^3  \int_{\partial \Omega_{t,i} ^\mathcal{H}} \tau \cdot \delta u^\mathcal{H}_i \, dx  \, .
 \end{equation}
Term $a_\sharp^\mathcal{H}$ is composed of terms that allows gluing the three domains together, using Nitsche's method. It is further expanded as
 \begin{equation}
a_\sharp^\mathcal{H}(u^\mathcal{H},\delta u^\mathcal{H}) = a_{1,\sharp}^\mathcal{H}(u^\mathcal{H},\delta u^\mathcal{H}) + a_{2,\sharp}^\mathcal{H}(u^\mathcal{H},\delta u^\mathcal{H}) +
a_{3,\sharp}^\mathcal{H}(u^\mathcal{H},\delta u^\mathcal{H}
)
\, .
\end{equation}
where the first term is for the matrix and the inclusions, while the second and third terms relate to the interface between coarse and fine domains. We have that
 \begin{equation}
 \begin{array}{rcl}
\displaystyle a_{i,\sharp}^\mathcal{H}(u^\mathcal{H},\delta u^\mathcal{H}) 
& = & \displaystyle \gamma_i \hat{w} _i
\int_{\Gamma^\mathcal{H}_i}  \llbracket u^\mathcal{H} \rrbracket_i \cdot \llbracket \delta u^\mathcal{H} \rrbracket_i \, dx
\\
& - & \displaystyle 
\int_{\Gamma^\mathcal{H}_i}  \left\{ t \right\}_i(u^\mathcal{H}) \cdot  
\llbracket \delta u^\mathcal{H} \rrbracket_i  \, dx
\\
& - & \displaystyle 
\int_{\Gamma^\mathcal{H}_i}  \left\{ t \right\}_i( \delta u^\mathcal{H}) \cdot  
\llbracket u^\mathcal{H} \rrbracket_i  \, dx \, ,
 \end{array}
\end{equation}
where
\begin{equation}
\begin{aligned}
\llbracket u^\mathcal{H} \rrbracket_1 = u_1^\mathcal{H} - u_2^\mathcal{H}\, , \\ 
\llbracket u^\mathcal{H} \rrbracket_2 = u_1^\mathcal{H} - u_3^\mathcal{H}\, ,  \\ 
\llbracket u^\mathcal{H} \rrbracket_3 = u_2^\mathcal{H} - u_3^\mathcal{H} \, ,
\end{aligned}
\end{equation}
denote the jumps in the displacements across $\Gamma^\mathcal{H}_1, \Gamma^\mathcal{H}_2$ and $\Gamma^\mathcal{H}_3$ respectively; and $\left\{ t \right\}_i$ denotes the following weighted averages 
\begin{equation}
\begin{aligned}
\left\{ t \right\}_1 = w^1_1 \sigma_1(\nabla_s u_1 ^\mathcal{H}) \cdot n_1 + w^2_1 \sigma_2(\nabla_s u_2 ^\mathcal{H}) \cdot n_1 \, ,\\
\left\{ t \right\}_2 = w^1_2 \sigma_1(\nabla_s u_1 ^\mathcal{H})\cdot n_2 + w^2_2 \sigma_3(\nabla_s u_3 ^\mathcal{H}) \cdot n_2 \, ,\\
\left\{ t \right\}_3 = w^1_3 \sigma_2(\nabla_s u_2 ^\mathcal{H}) \cdot n_2 + w^2_3 \sigma_3(\nabla_s u_3 ^\mathcal{H}) \cdot n_2\, ,
\end{aligned}
\end{equation}
where $n_1 = - \frac{\nabla \phi_1}{| \nabla \phi_1 |}$, $n_2 = - \frac{\nabla \phi_2}{| \nabla \phi_2 |}$.
\begin{equation}
\begin{aligned}
w^1_1 = \frac{E_2}{E_1 + E_2 }, \quad w^2_1 = \frac{E_1}{E_1 + E_2} \\
w^1_2 = \frac{\frac{E_3}{H}}{\frac{E_1}{h} + \frac{E_3}{H}}, \quad w^2_2 = \frac{\frac{E_1}{h}}{\frac{E_1}{h} + \frac{E_3}{H}} \\
w^1_3 = \frac{\frac{E_3}{H}}{\frac{E_2}{h} + \frac{E_3}{H}}, \quad w^2_3 = \frac{\frac{E_2}{h}}{\frac{E_2}{h} + \frac{E_3}{H}} \, .
\end{aligned}
\end{equation}

\begin{equation}
\begin{aligned}
\hat{w} _1= \frac{E_1 E_2}{h (E_1 + E_2) },\\
\hat{w} _2 = \frac{\frac{E_1}{H}\frac{E_3}{h}}{\frac{E_1}{h} + \frac{E_3}{H}}, \\
\hat{w} _3 = \frac{\frac{E_2}{H} \frac{E_3}{h}}{\frac{E_2}{h} + \frac{E_3}{H}} \, .
\end{aligned}
\end{equation}


Finally, $a_\heartsuit ^\mathcal{H}$ is an interior penalty regularisation term that reads as, for $i \in \{ 1, 2, 3\}$,
\begin{equation}
a_\heartsuit ^\mathcal{H}(u^\mathcal{H},\delta u^\mathcal{H})  =
\sum_{F\in {\hat{\mathcal{F}}}_i ^G} \bigg(\int_F {\beta_i \mathcal{H}_i}  \llbracket {\nabla_s{u^\mathcal{H}}} \rrbracket \llbracket {\nabla_s{(\delta u^\mathcal{H})}} \rrbracket \ dx \bigg)
\, ,
\end{equation}
where $\beta_i > 0$ is the ghost penalty parameter, and $\llbracket x \rrbracket$ is the normal jump of quantity $x$ over the $F$. As explained in \cite{BURMAN2012328, Burman.Claus.ea.15}, the ghost penalty term extends the coercivity from the physical domain into the discretised domain. 

\section{Numerical results}
\label{Section3}

In this section, we first verify the proposed multiresolution framework for a simplified multiscale elasticity problem. Then, we adopt von Mises material for the multiscale model and assess it for two types of hard and void micro-inclusions. Eventually, we assess the performance of the zoom technique for the zooms with time-dependent geometrical properties. All the numerical results are produced by the CutFEM library \cite{Burman.Claus.ea.15}, developed in FEniCS \cite{fenics}.

\subsection{Verification test: Quasi-uniform porous structure}
\begin{figure}[h]
\centering
  \includegraphics[scale=.27]{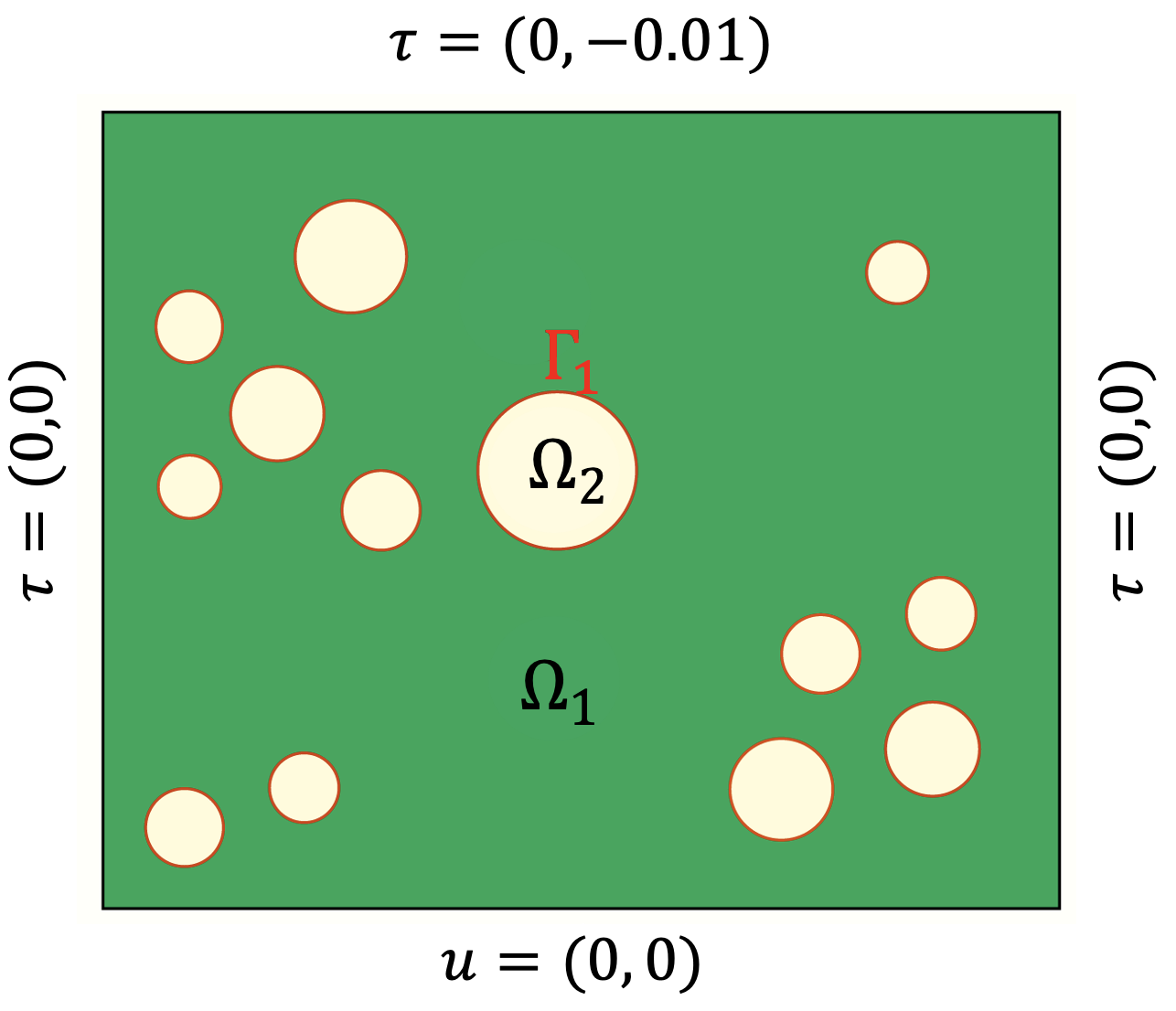}
  \caption{Boundary conditions and geometry of a heterogeneous structure with compression test}
  \label{fig:model1schm}
\end{figure}

Here, the proposed multiresolution CutFEM framework is assessed for a heterogeneous structure with micropores and then compared with the corresponding references (i.e. a full microscale FEM and the smoothed concurrent multiscale method that we proposed in \cite{mikaeili2022concurrent}). We consider the same quasi-uniform porous medium given in \cite{mikaeili2022concurrent} which includes circular pores distributed all over the domain (as depicted in Figure \ref{fig:model1schm}). The material behaviour is assumed as elastic and isotropic. According to \cite{mikaeili2022concurrent}, the material properties for matrix are given as $E_1=1$ and $\nu_1 = 0.3$, and for the homogenised model are derived by Mori-Tanaka (MT) method \cite{Mura87, IMANI201816489} as following: $E_3=0.78$ and $\nu_3 = 0.3$. 

\begin{figure}[h]
   \centering
     \subfloat[ \label{FEMmesh1}]{%
      \includegraphics[scale=0.27]{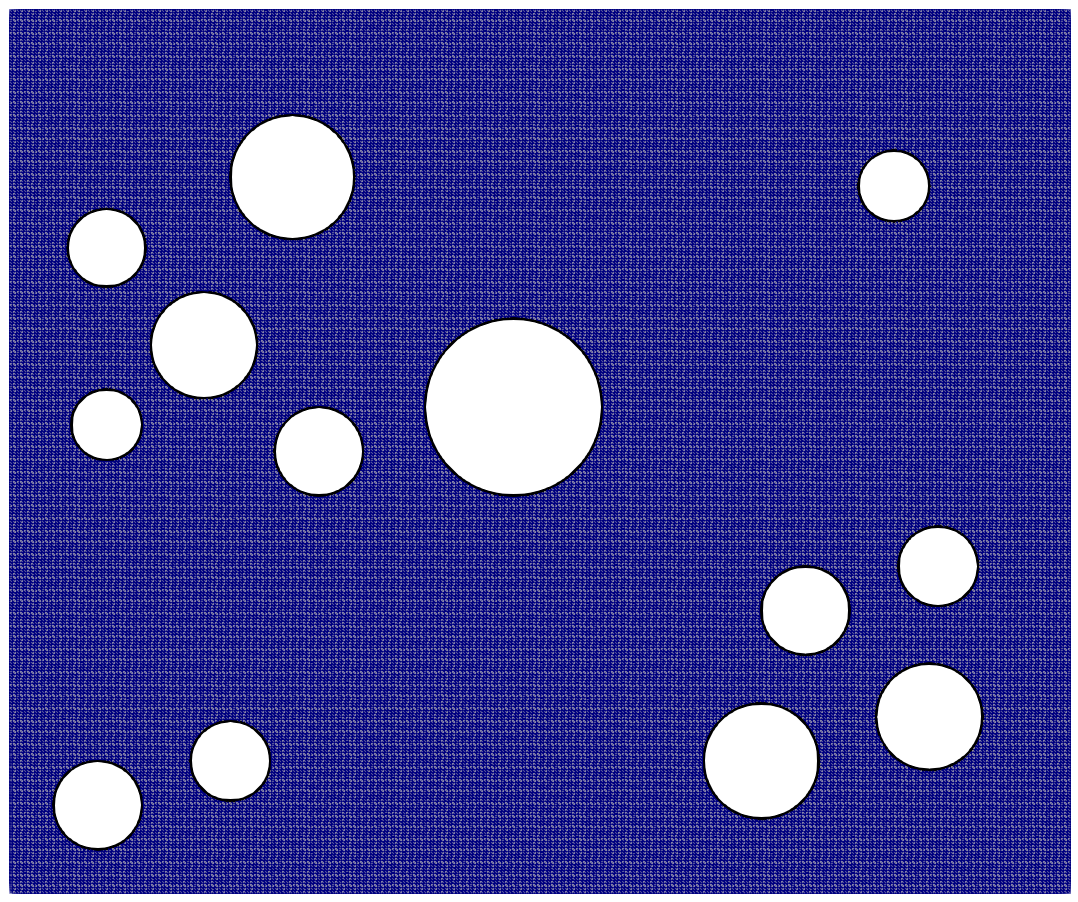}
     }
    \hfill
    \subfloat[ \label{MultiMesh1}]{%
     \includegraphics[scale=0.15]{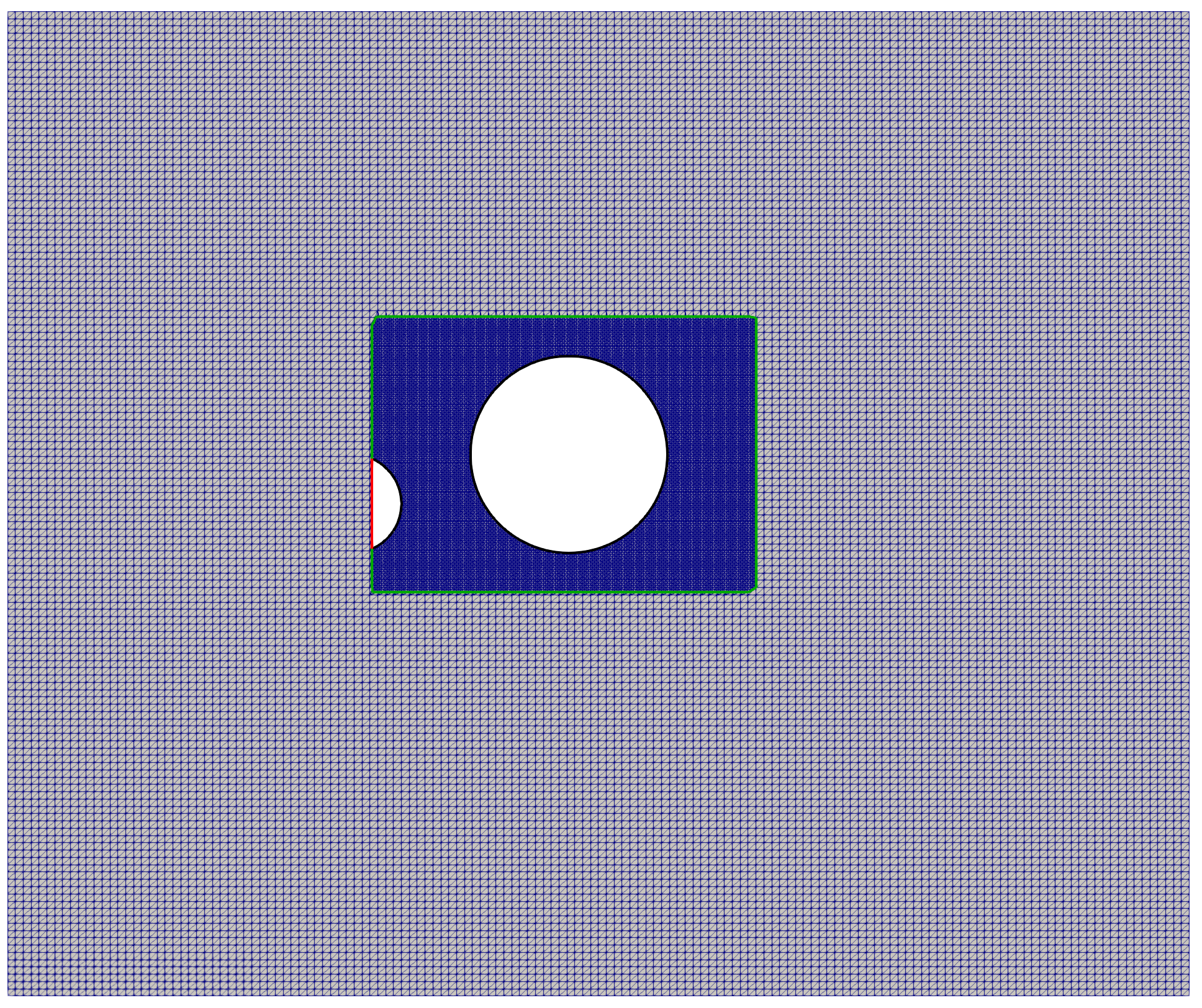}
     }
    \hfill
     \caption{Discretised domains; a) FE conforming mesh and b) CutFEM non-conforming multiresolution mesh.}
     \label{fig:Model1meshesS}
\end{figure}

\begin{figure}[h]
\centering
  \includegraphics[scale=.24]{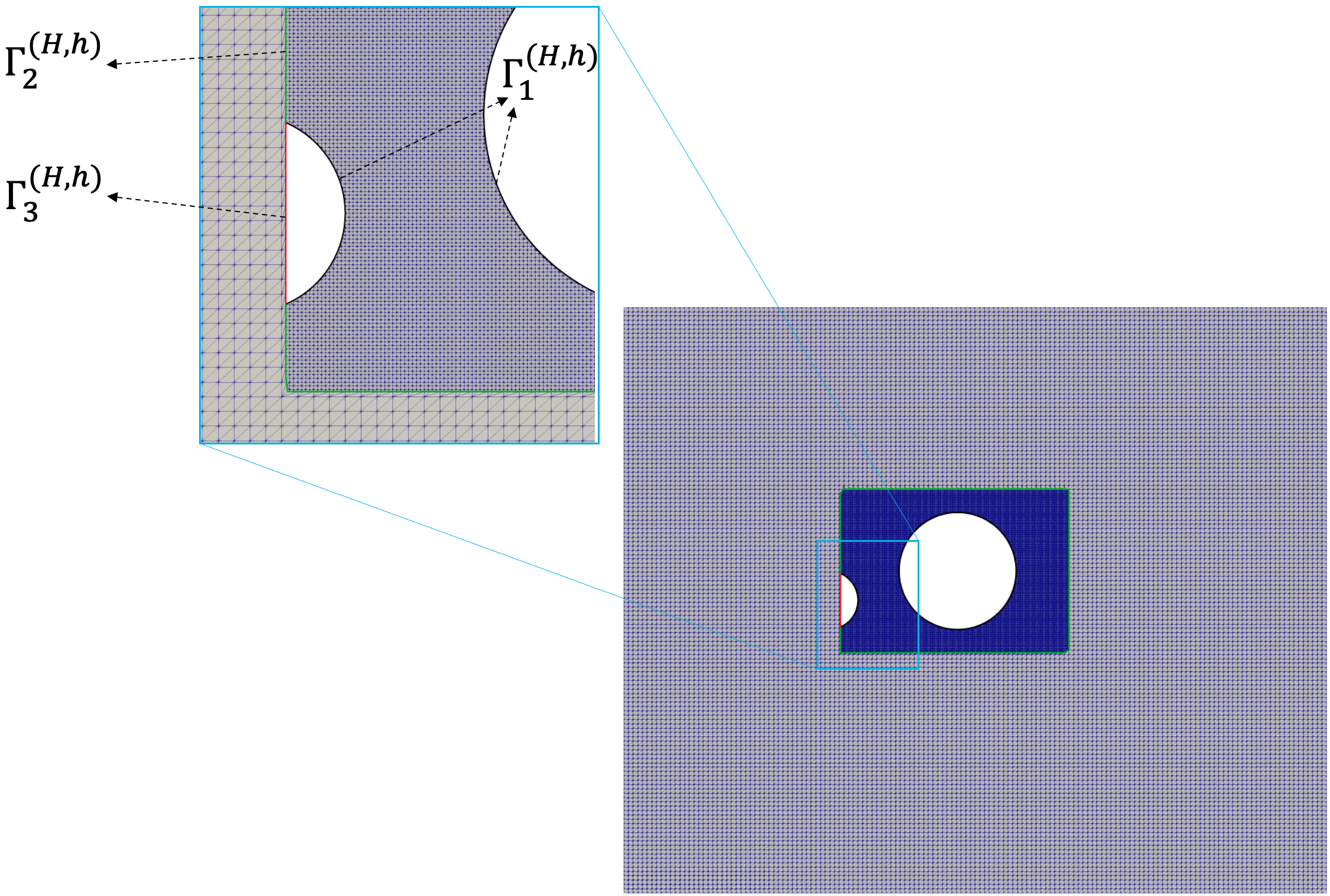}
  \caption{Discretised domain for the multiscale CutFEM including interfaces $\Gamma_1 ^{(H,h)}$, $\Gamma_2 ^{(H,h)}$ and $\Gamma_3 ^{(H,h)}$. The cut elements intersected by $\Gamma_1 ^{(H,h)}$, $\Gamma_2 ^{(H,h)}$ and $\Gamma_3 ^{(H,h)}$ are shown with their integration subtriangles.}
  \label{fig:model1meshzoom}
\end{figure}

The computational meshes for the full microscale FEM and multiresolution CutFEM models are shown in Figure \ref{fig:Model1meshesS}(a) and (b), consisting of linear Lagrangian elements with a smallest element size of $h_{\text{min}}=0.054$. The element size within the zoom area (referred to as $h$) matches that of the reference models for verification purposes, while in the coarse region it is set to $H=0.11$. As depicted in Figure \ref{fig:model1meshzoom}, the three interface types, $\Gamma_{1} ^{(H,h)}$, $\Gamma_2 ^{(H,h)}$, and $\Gamma_3 ^{(H,h)}$, intersect the background mesh in arbitrary ways. The ghost penalty parameters for the intersected elements are specified as $\beta_1 = \beta_2 = \beta_3 = 0.005$. Additionally, the Nitsche's penalty parameters used to couple the micro and macro-scale models are $\gamma_2 = \gamma_3 = 10$.


\begin{figure}[h]
   \centering
     \subfloat[ \label{DisFEM1}]{%
      \includegraphics[scale=0.2]{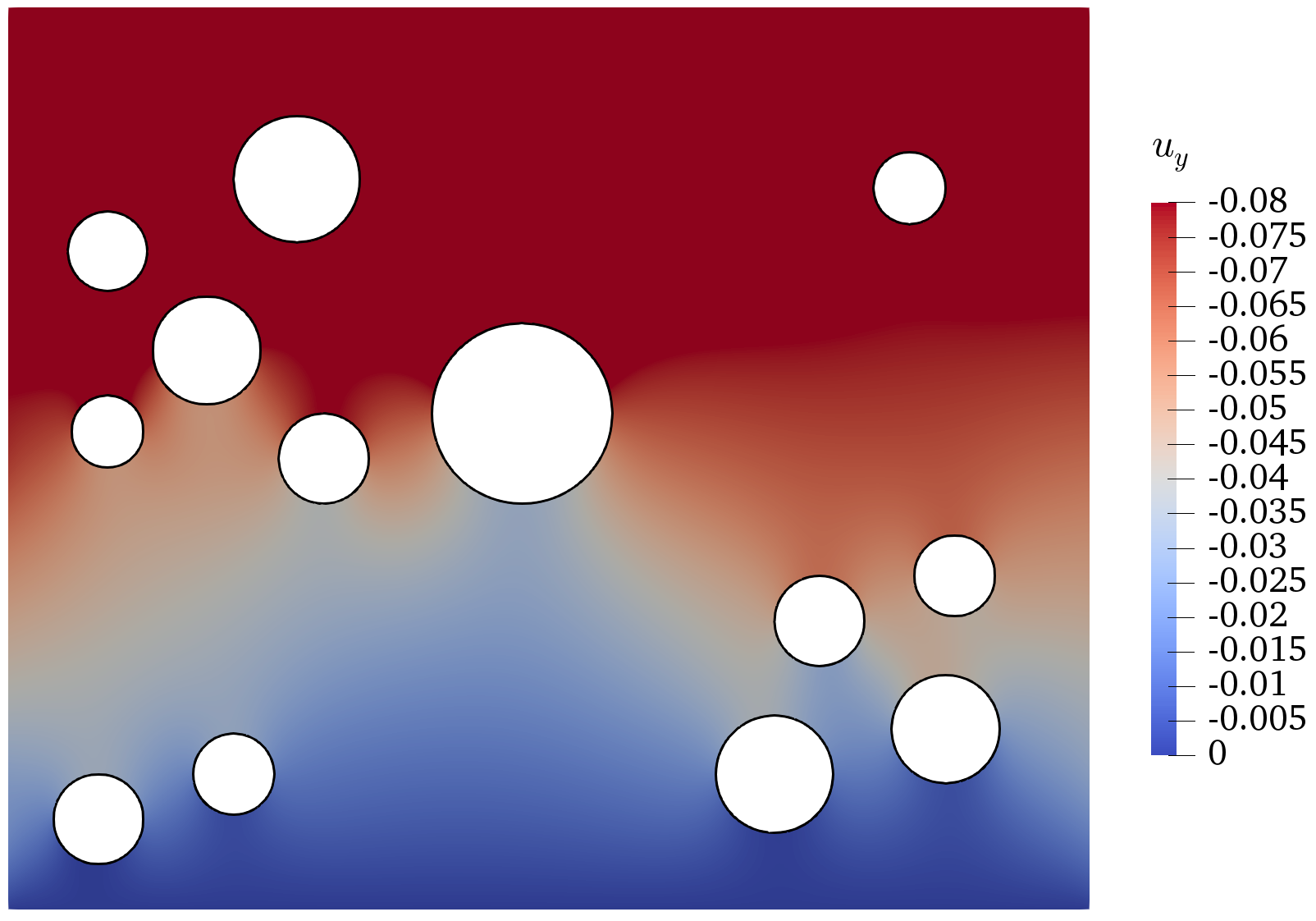}
     }
    \hfill
    \subfloat[ \label{DisMulti1}]{%
     \includegraphics[scale=0.21]{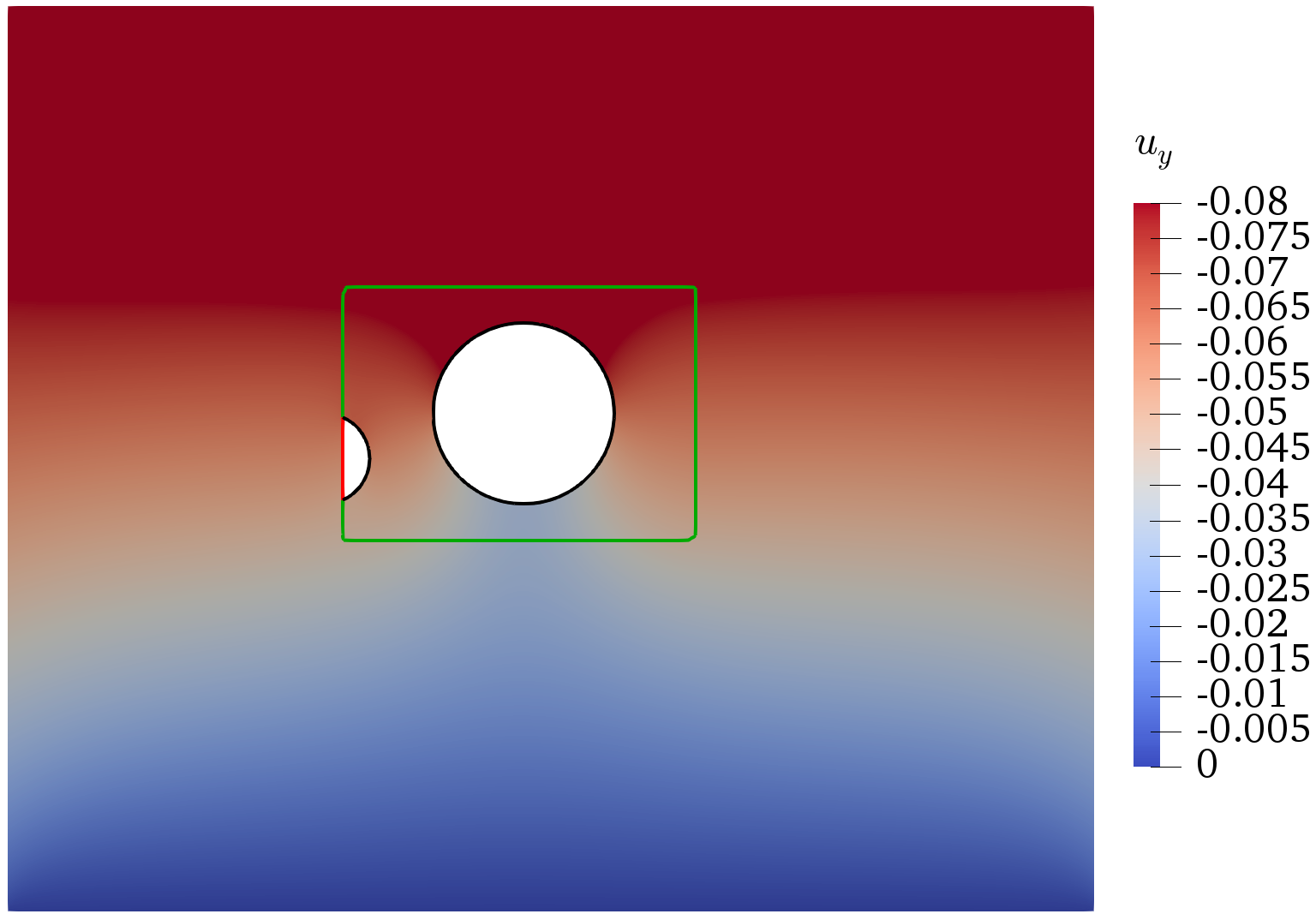}
     }
    \hfill
     \caption{Displacement field component $u_y$; a) microscale FEM and b) multiscale CutFEM.}
     \label{fig:Model1U}
\end{figure}

A compression test is conducted for the heterogeneous structure where the displacements are fixed along the $x$-direction and $y$-direction on the lower end, and traction $\tau =(0, -0.01)$ is prescribed along the top edge. The FEM displacement component $u_y$ contour is obtained and used as a reference solution, see Figure \ref{fig:Model1U} (a). The same test is carried out for the multiscale model. The corresponding displacement field component is shown in Figure \ref{fig:Model1U} (b). When our multiscale model is compared with the full microsale FEM and the smoothed multiscale models, a close similarity of $u_y$ is observed inside the zooming region. Outside of the zoom, again, a satisfactory agreement is achieved.

\begin{figure}[h]
   \centering
     \subfloat[ \label{SigFEM1}]{%
      \includegraphics[scale=0.25]{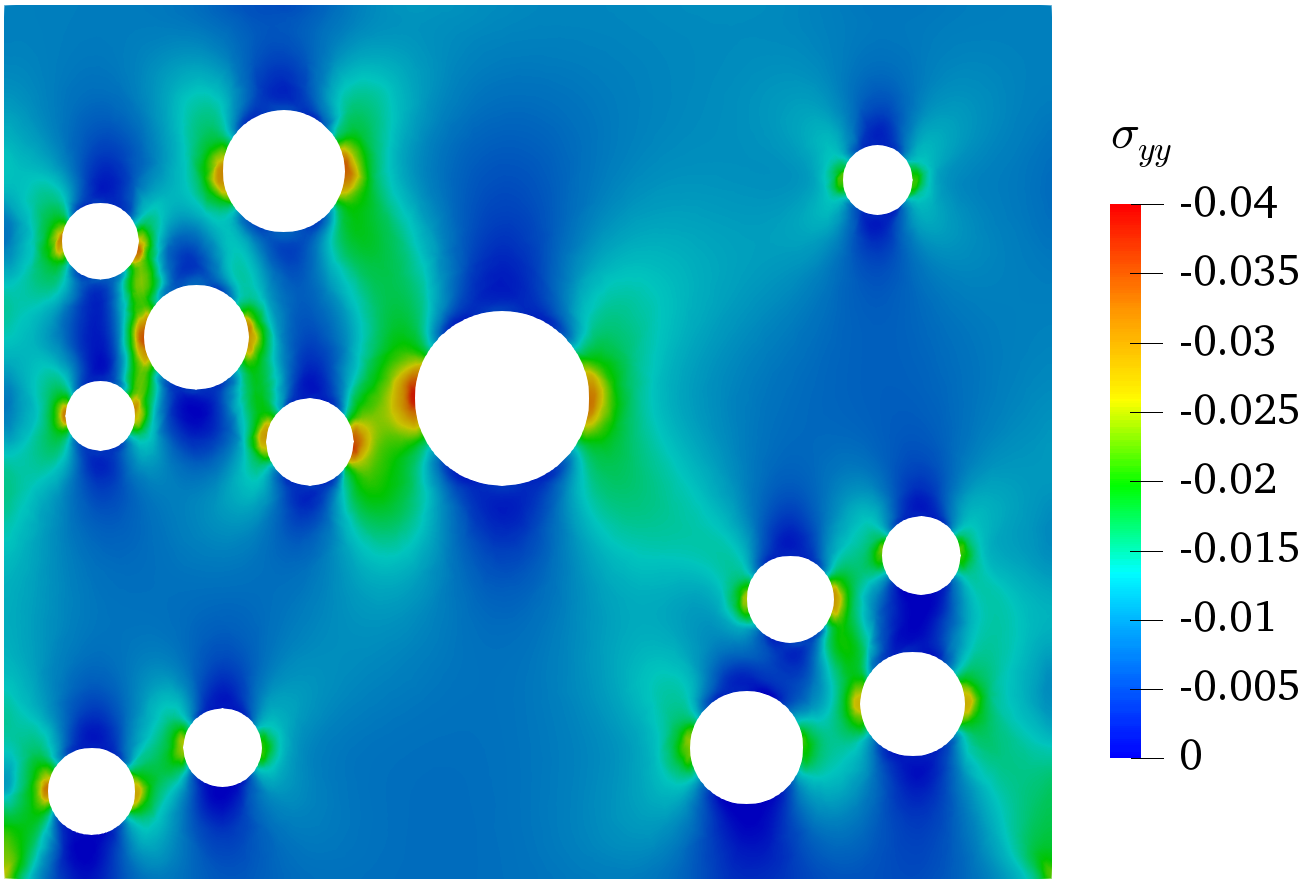}
     }
    \hfill
    \subfloat[ \label{SigMulti1}]{%
     \includegraphics[scale=0.25]{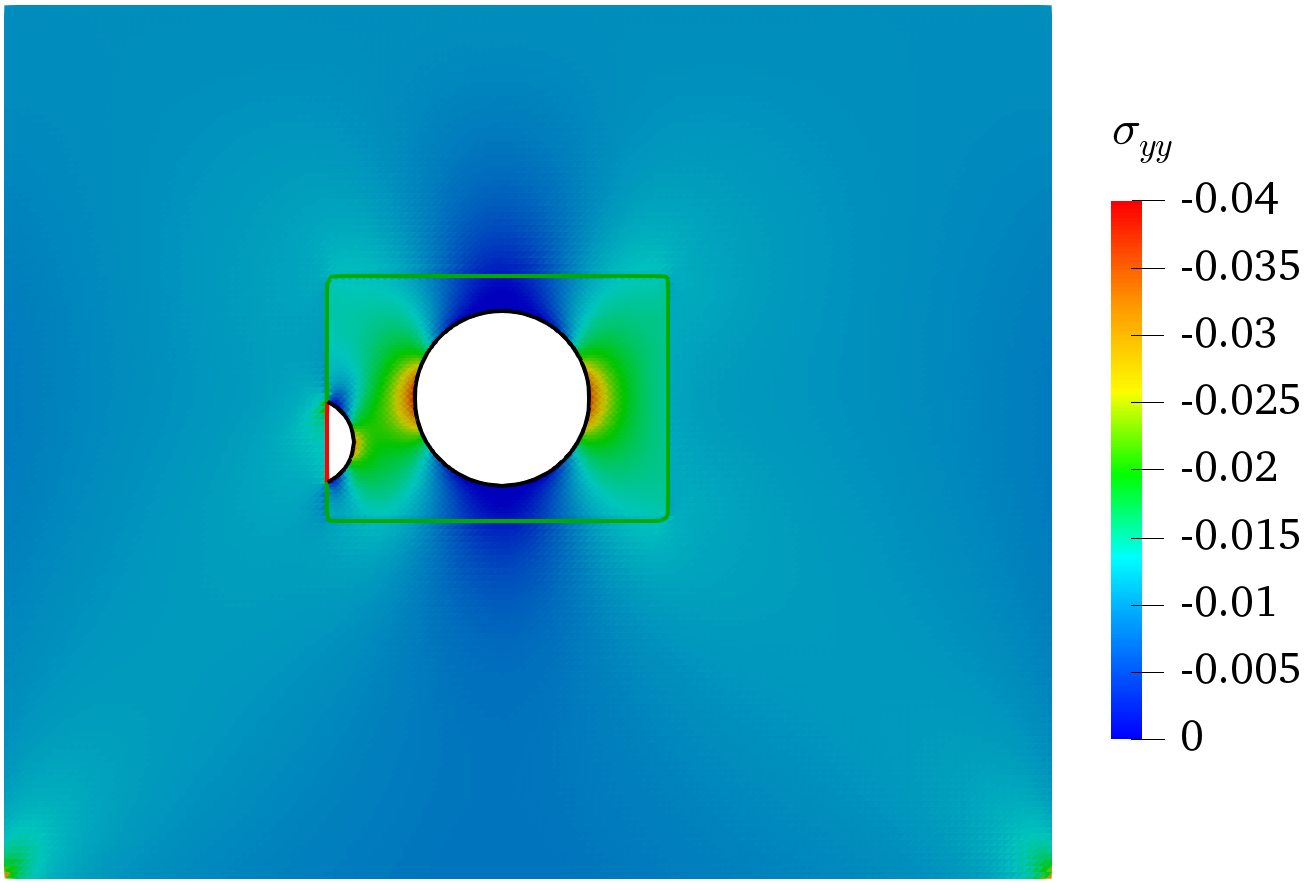}
     }
    \hfill
     \caption{Stress field component $\sigma_{yy}$; a) microscale FEM and b) multiscale CutFEM.}
     \label{fig:Model1sigma}
\end{figure}

We compute the stress field component $\sigma_{yy}$ in Figure \ref{fig:Model1sigma}. Here again, a good agreement is achieved between the multiscale and reference models. The homogeneous model adopted in the coarse domain of the multiscale model smooth out the fluctuations produced by the coarse pores, and the overall trend in this domain is captured very well.

\begin{figure}[h]
   \centering
     \subfloat[ \label{errormesh}]{%
      \includegraphics[scale=0.33]{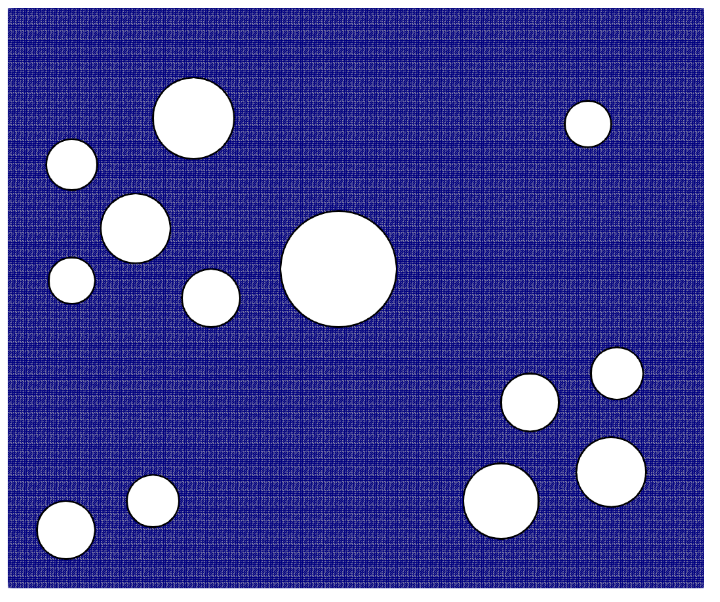}
     }
    \hfill
    \subfloat[ \label{coarsemulti}]{%
     \includegraphics[scale=0.12]{Model1meshCutB.png}
     }
    \hfill
        \subfloat[ \label{coarsemulti}]{%
     \includegraphics[scale=0.12]{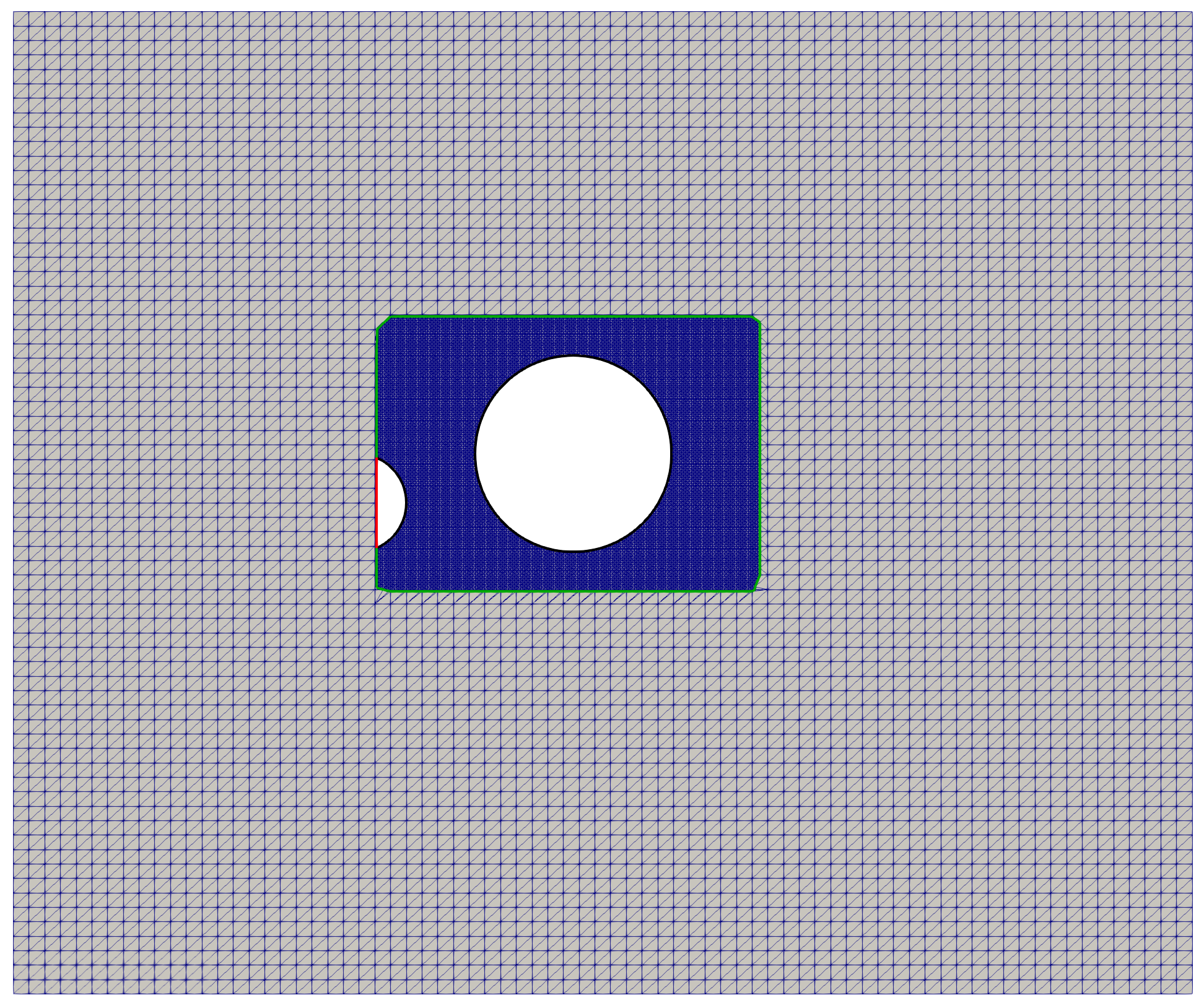}
     }
    \hfill
     \caption{Computational meshes for; a) computing the error field ($h_{min}=0.054$), b) multiresolution CutFEM with a mesh size $H=0.11$ in the macroscale region and c) multiresolution CutFEM with a mesh size $H=0.22$ in the macroscale region. For both multiresolution CutFEM models we have $h = 0.054$.
     All the cut elements in the multiresolution CutFEM are shown with their integration subtriangles.}
     \label{fig:Model1errormesh}
\end{figure}

 For further investigations, we study the effects of mesh coarsening in the coarse region of the multiresolution framework on the energy norm of the error field. The corresponding mesh layouts for two types of multiresolution models are depicted in Figures \ref{fig:Model1meshesS}(b) and (c). While the mesh outside the zoom differs for the two multiresolution models, the mesh inside is considered to be the same size ($h=0.054$). Moreover, as shown in Figure \ref{fig:Model1errormesh} (a), a fully fine resolution mesh is used for computing the error field with $h_{\text{min}}=h=0.054$. We compute the energy norm of the error field with respect to the reference FE model using the following formulation:
 
\begin{equation}
\| e \| =  \sqrt{  \int_{\Omega} \nabla_s e : \nabla_s e \, dx \, }. 
\end{equation}
where $e = u_{\text{ref}} - u$. Here, $u_{\text{ref}}$ and $u$ denote the displacement for the FE and multiresolution models, respectively.

\begin{figure}[h]
   \centering
     \subfloat[ \label{EnormMultiA}]{%
      \includegraphics[scale=.9]{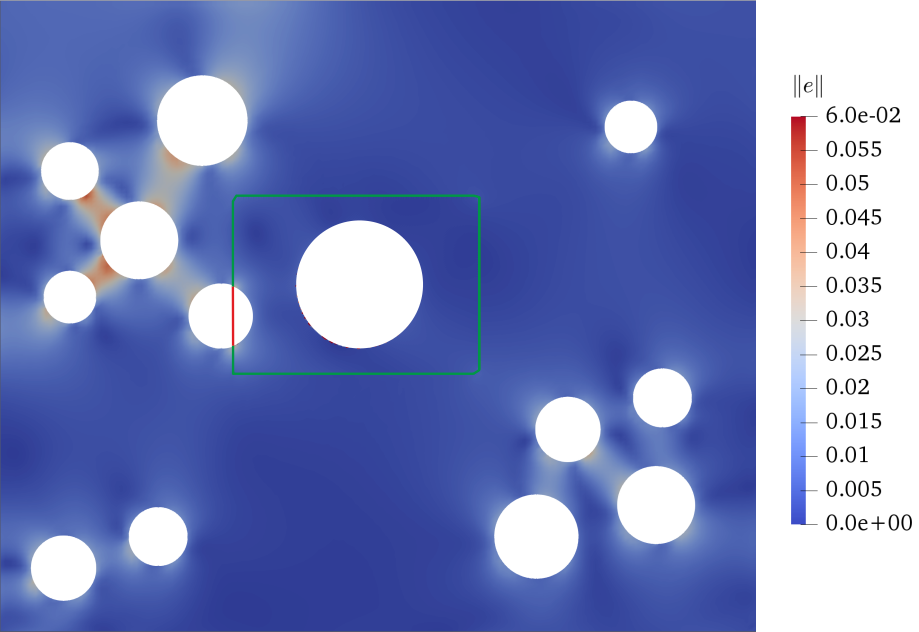}
     }
    \hfill
    \subfloat[ \label{EnormMultiB}]{%
     \includegraphics[height = 48.4mm, width=70mm, scale=.9]{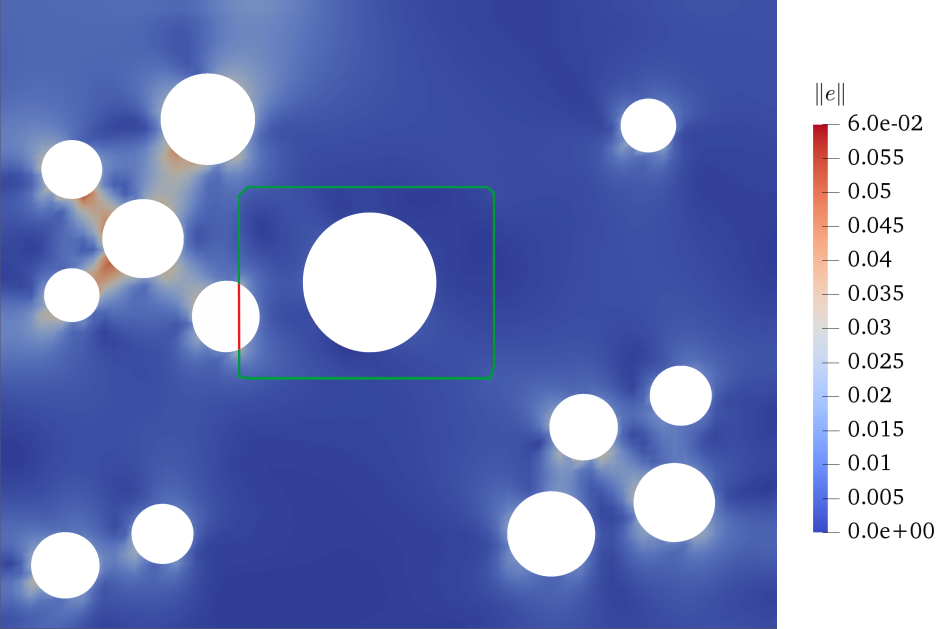}
     }
    \hfill
     \caption{Energy norm of error field $\| e \|$ for multiscale CutFEM with different mesh resolutions in the macroscale region; a) $H=0.11$ and b) $H=0.22$. A uniform mesh size of $h=0.054$ is used for plotting the $\| e \|$.}
     \label{fig:Model1error}
\end{figure}

The energy norm of the error fields for the two multiresolution models is computed as shown in Figure \ref{fig:Model1error}, where $H=0.11$ for the first multiresolution model and $H=0.22$ for the second one. The results indicate that the error norm within the zoomed region remains minimal for both models, irrespective of the coarsening of the mesh outside the zoomed area.
Furthermore, the analysis suggests that the primary source of error outside the zoomed region can be attributed to the homogenisation approximation rather than the mesh size. This assertion is supported by the observation that the $\lVert e \rVert$ remains unaltered in the coarse region despite the mesh coarsening process.


\clearpage

\subsection{S shape heterogeneous structure with a stationary zoom}
\label{Results:fixedSshape}
\begin{figure}[h]
   \centering
     \subfloat[ \label{SchS1}]{%
      \includegraphics[scale=0.32]{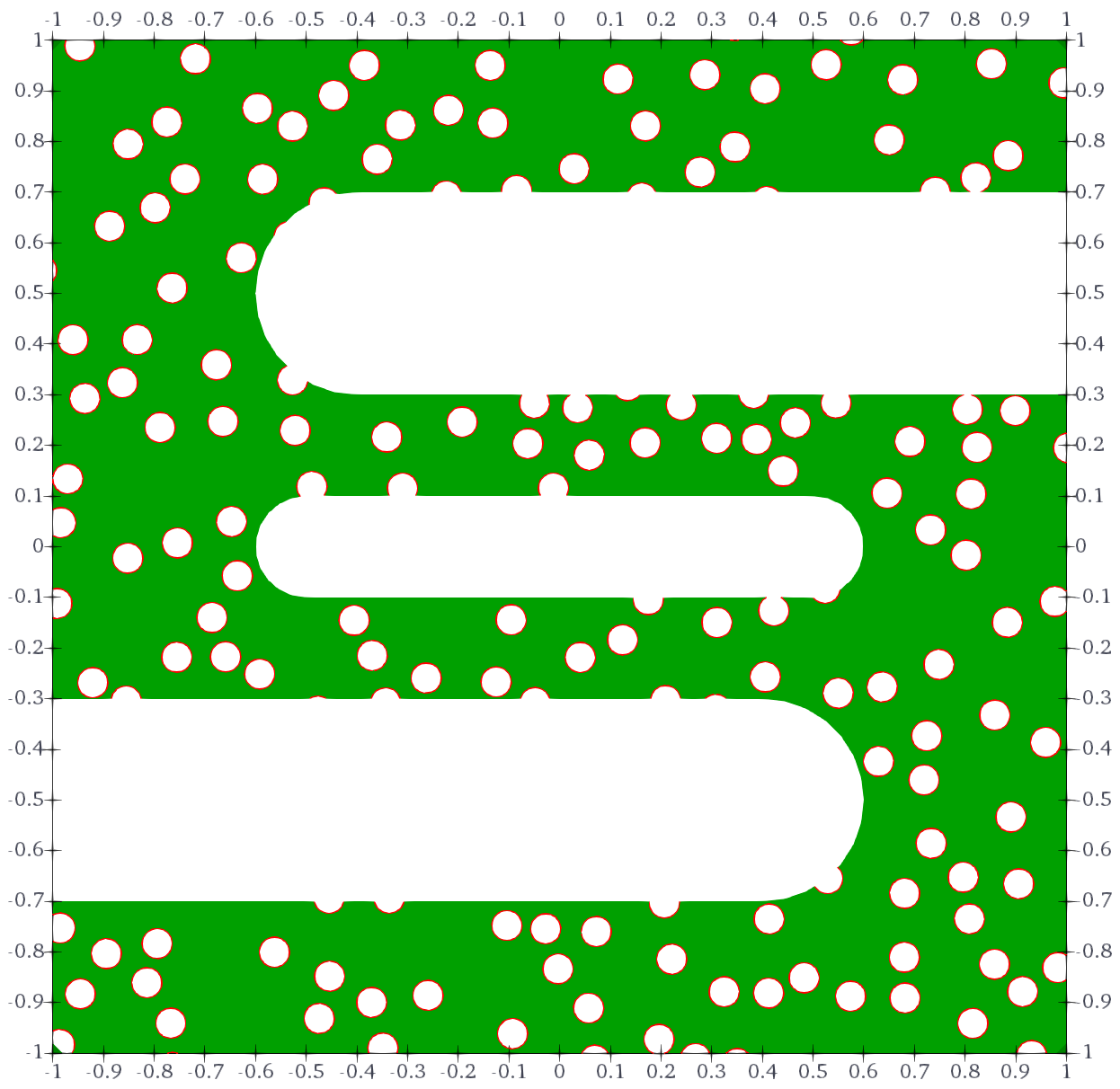}
     }
    \hfill
    \subfloat[ \label{SchS2}]{%
     \includegraphics[scale=0.32]{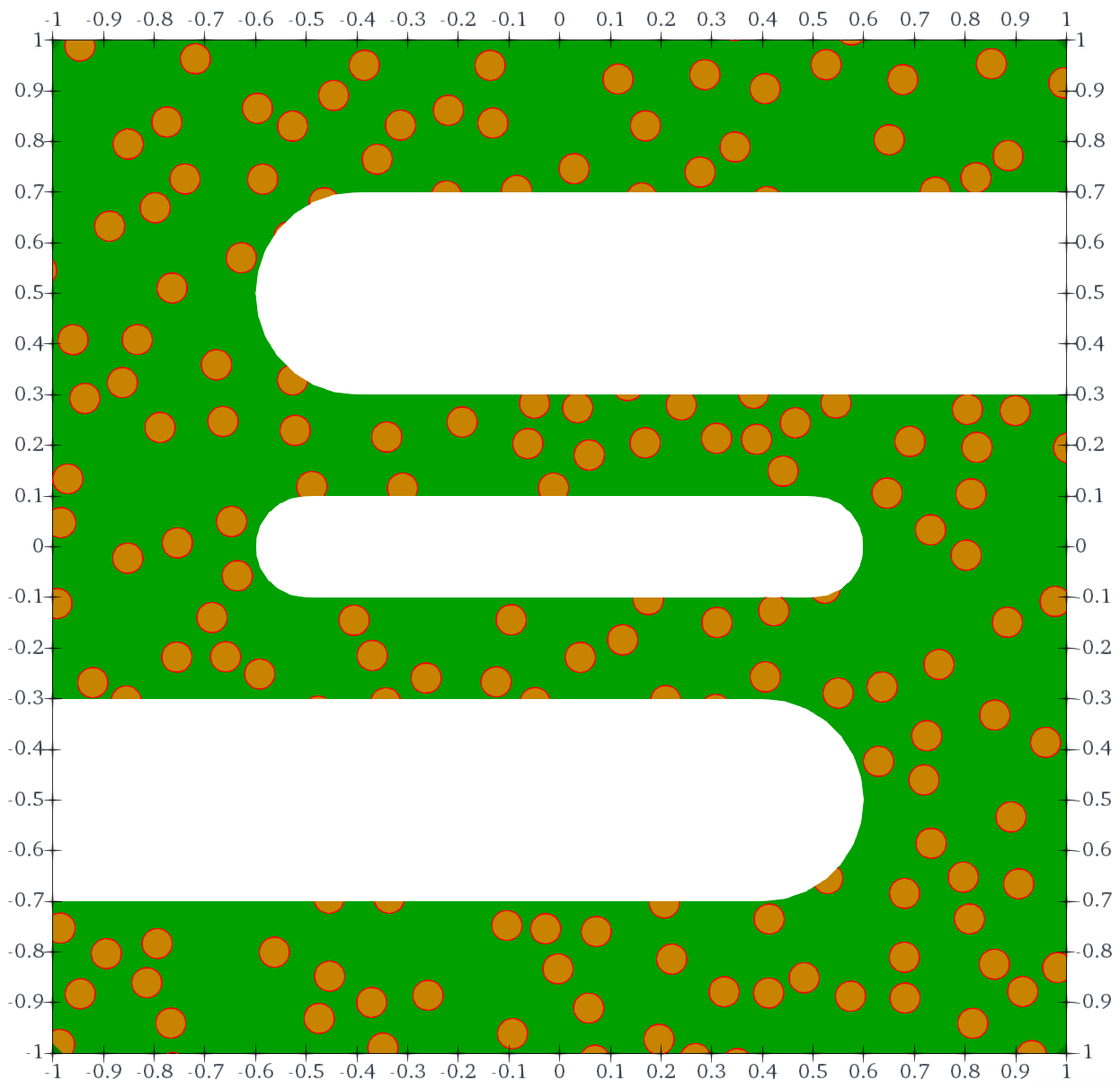}
     }
    \hfill
     \caption{Geometry of heterogeneous structures; a) heterogeneities are voids, b) heterogeneities are hard inclusions.}
     \label{fig:model2schmincl}
\end{figure}

In this section, we assess the ability and efficiency of the multiresolution CutFEM in modelling heterogeneous structures with nonlinear material properties and different types of heterogeneities. We consider an S shape heterogeneous structure with a random distribution of heterogeneities. As shown in Figure \ref{fig:model2schmincl}, the heterogeneities can be either voids or hard inclusions. We assume von Mises elastoplastic material behaviour for these structures. The material properties heterogeneous structures are: $2E_1=E_2=2$, $\nu_1=\nu_2=\nu_3=0.3$, ${Y}_{0,1}= {Y}_{0,2} = {Y}_{0,3} =0.25$ and $\hat{H}^p _1= \hat{H}^p _2= \hat{H}^p _3= 10^{-2}$. The material properties for the macroscale homogenised model with voids and inclusions are calculated by using MT as follows, respectively: $E_3= 0.5, 1.3$. To analyse the influence of different microstructural features on the accuracy of the proposed multiscale framework, we consider the geometry and distribution of the voids and hard inclusions to be similar in the two structures. We restrict the displacement along the $x$ and $y$ directions on the lower end and apply traction $\tau =(0, 0.18)$ incrementally along the top edge of the structure.

\begin{figure}[h]
   \centering
        \subfloat[ \label{Coarse}]{%
      \includegraphics[scale=0.223]{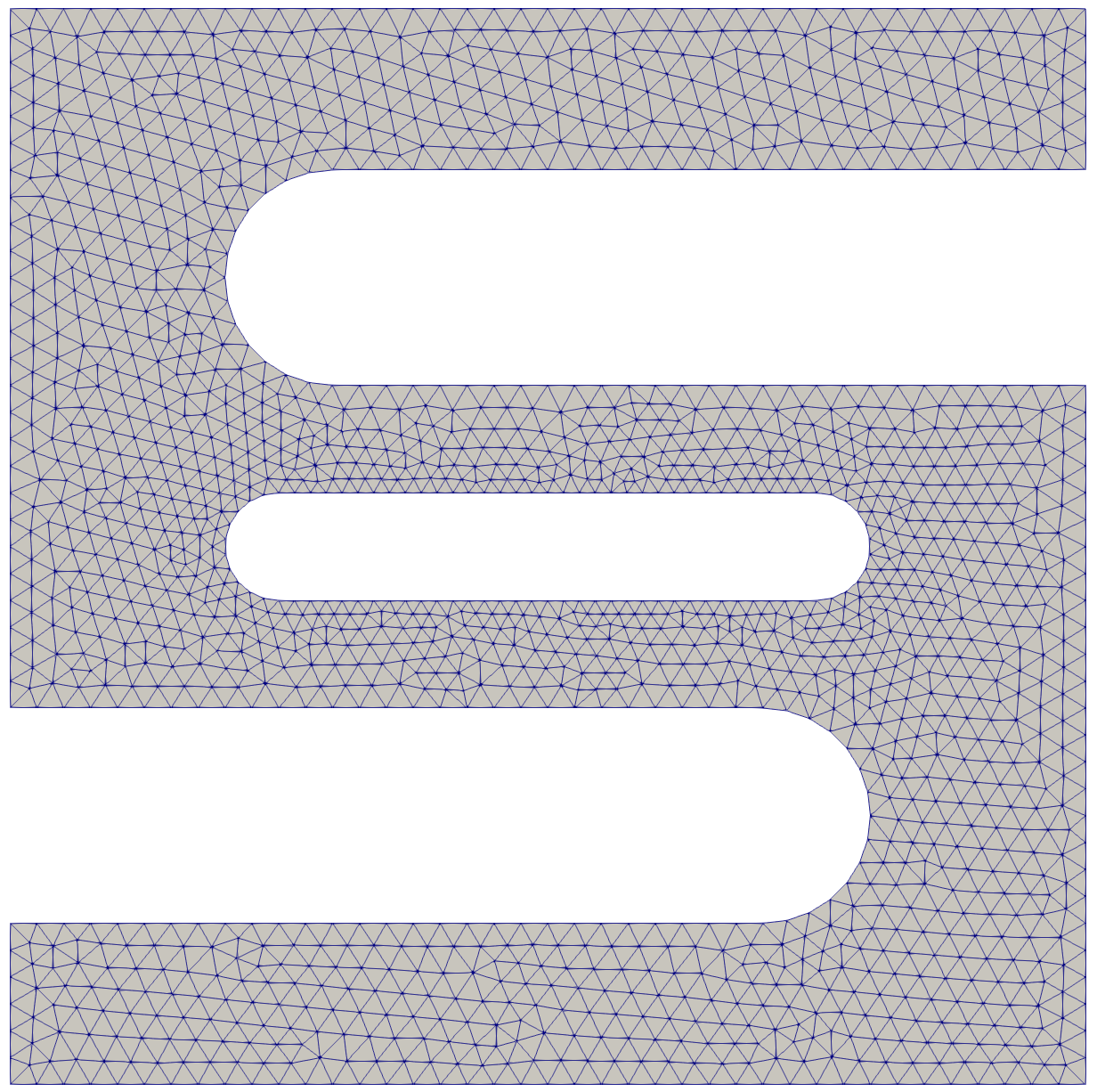}
     }
    \hfill
     \subfloat[ \label{Pores}]{%
      \includegraphics[scale=0.27]{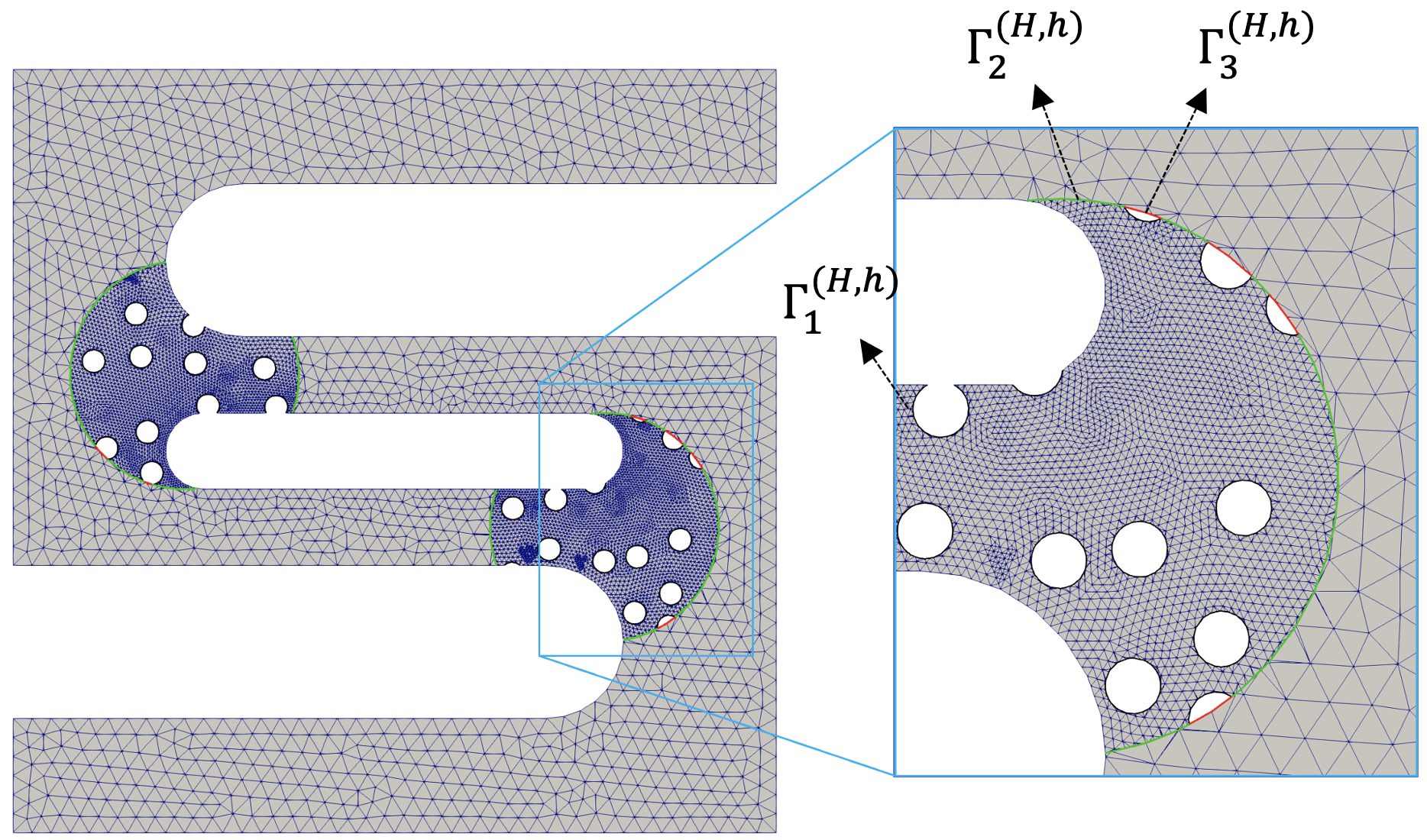}
     }
    \hfill
    \subfloat[ \label{HardInc}]{%
  \includegraphics[scale=.27]{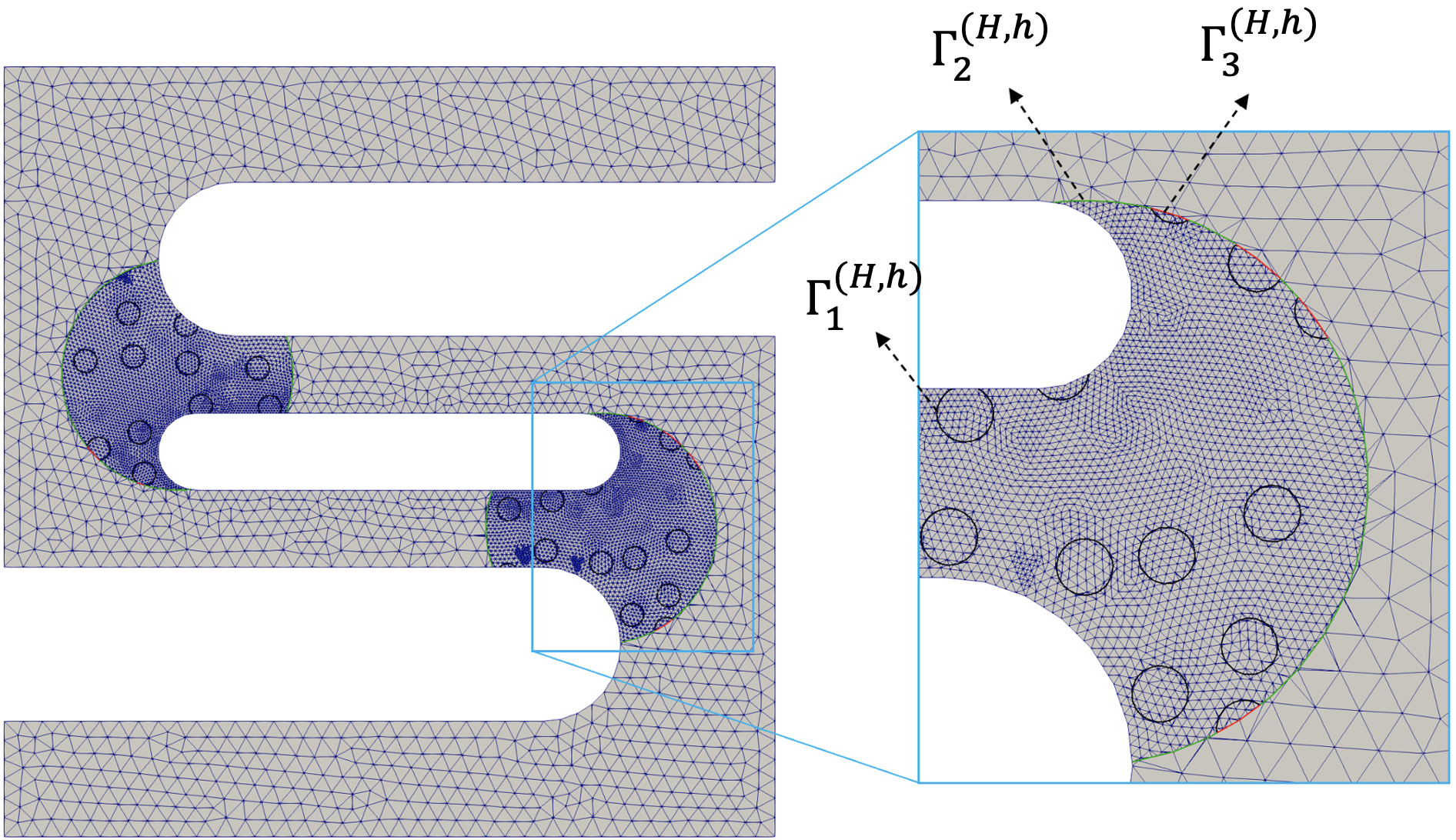}
     }
    \hfill
         \caption{Computational meshes; a) coarse mesh, b) multiresolution mesh for the porous microstructure, c) multiresolution mesh for the microstructure with hard inclusions. The cut elements are depicted with their integration subtriangles. }
  \label{fig:model2meshzoom}
\end{figure}

We employ two circular zooms fixed over a background mesh (see Figure \ref{fig:model2meshzoom}). We refine the mesh inside the zoom regions with a refinement scale defined as $s=\frac{1}{16}$ (means each coarse element is subdivided hierarchically into 16 fine elements), where the largest element size in the macroscale region is $H=0.06$. The discretised physical domain of multiresolution models in Figures \ref{fig:model2meshzoom}(b) and (c) show that all the three interfaces intersect the coarse background mesh (see Figure \ref{fig:model2meshzoom} (a)) for both models in a fully arbitrary manner. For these models, the ghost penalty regularisation and Nitsche's terms are set as: $\beta_1 = \beta_2 = \beta_3 = 1$ and $\gamma_1 = \gamma_2 = \gamma_3 = 10$.

\begin{figure}[h]
   \centering
     \subfloat[ \label{DisMulti2A}]{%
      \includegraphics[scale=0.32]{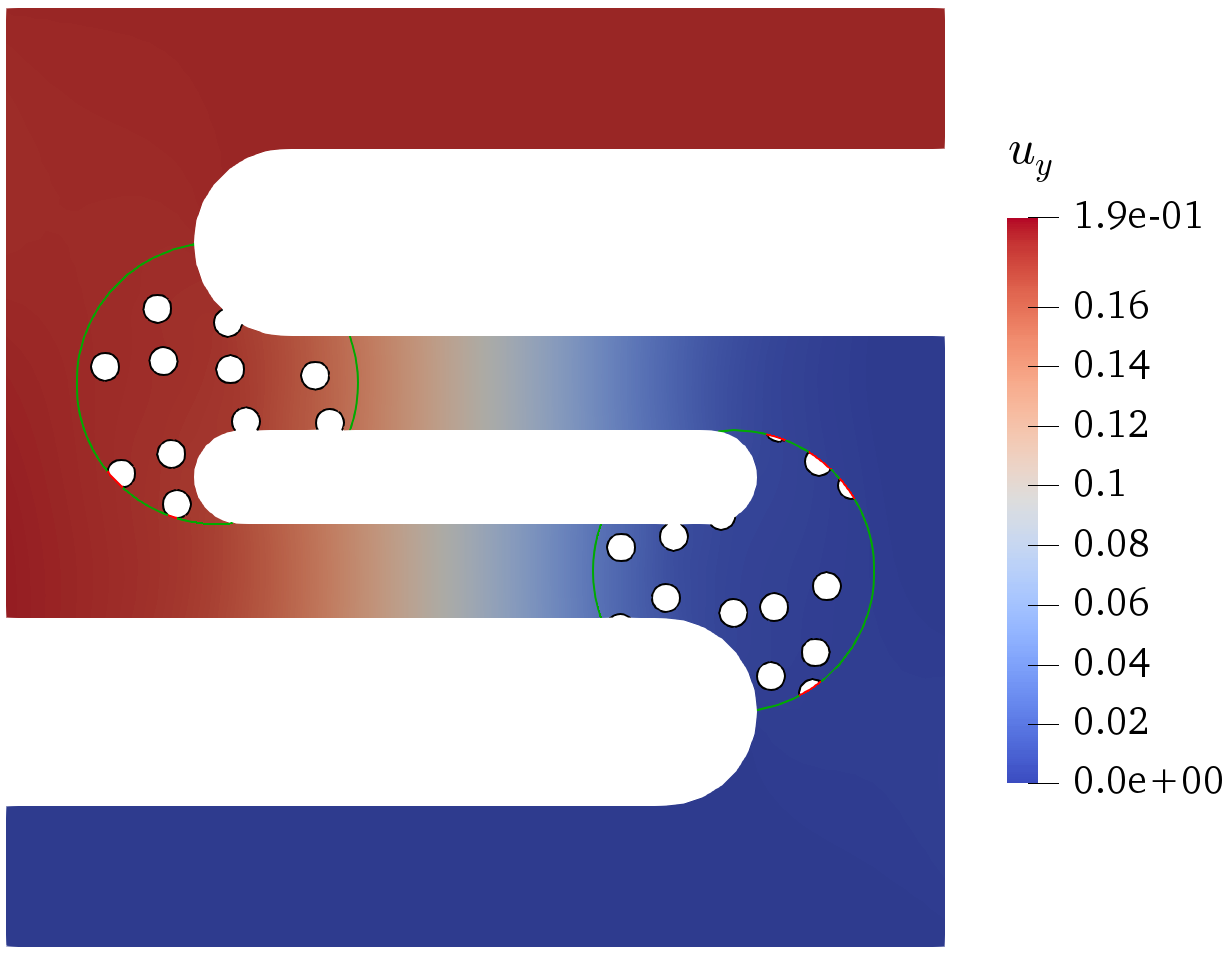}
     }
    \hfill
    \subfloat[ \label{DisMulti2B}]{%
     \includegraphics[scale=0.32]{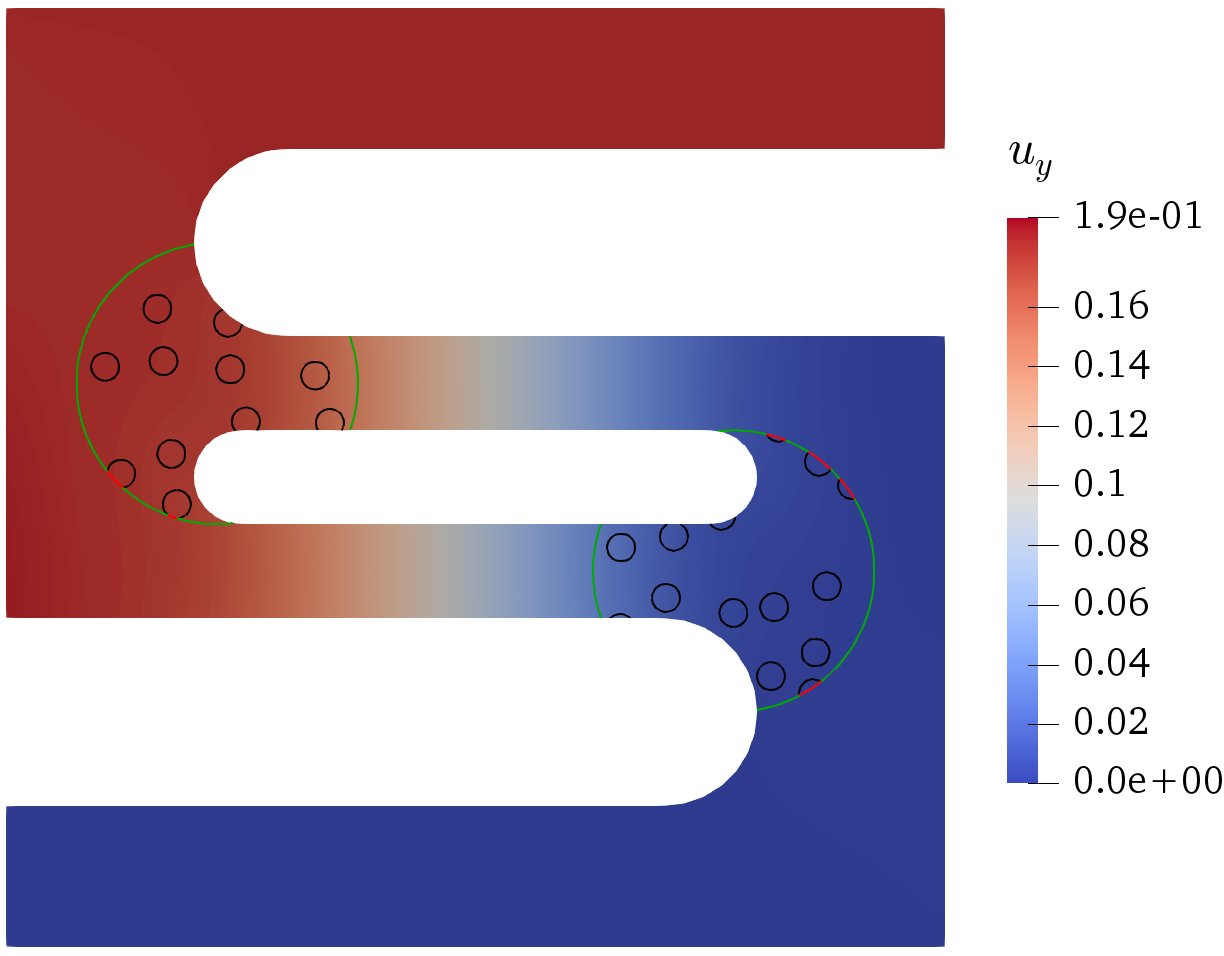}
     }
    \hfill
     \caption{Displacement field $u_y$ for the heterogeneous structures in the last time step; a) heterogeneities are voids, b) heterogeneities are hard inclusions}
     \label{fig:model2disp}
\end{figure}

\begin{figure}[h]
   \centering
     \subfloat[ \label{SigMultiA}]{%
      \includegraphics[scale=0.32]{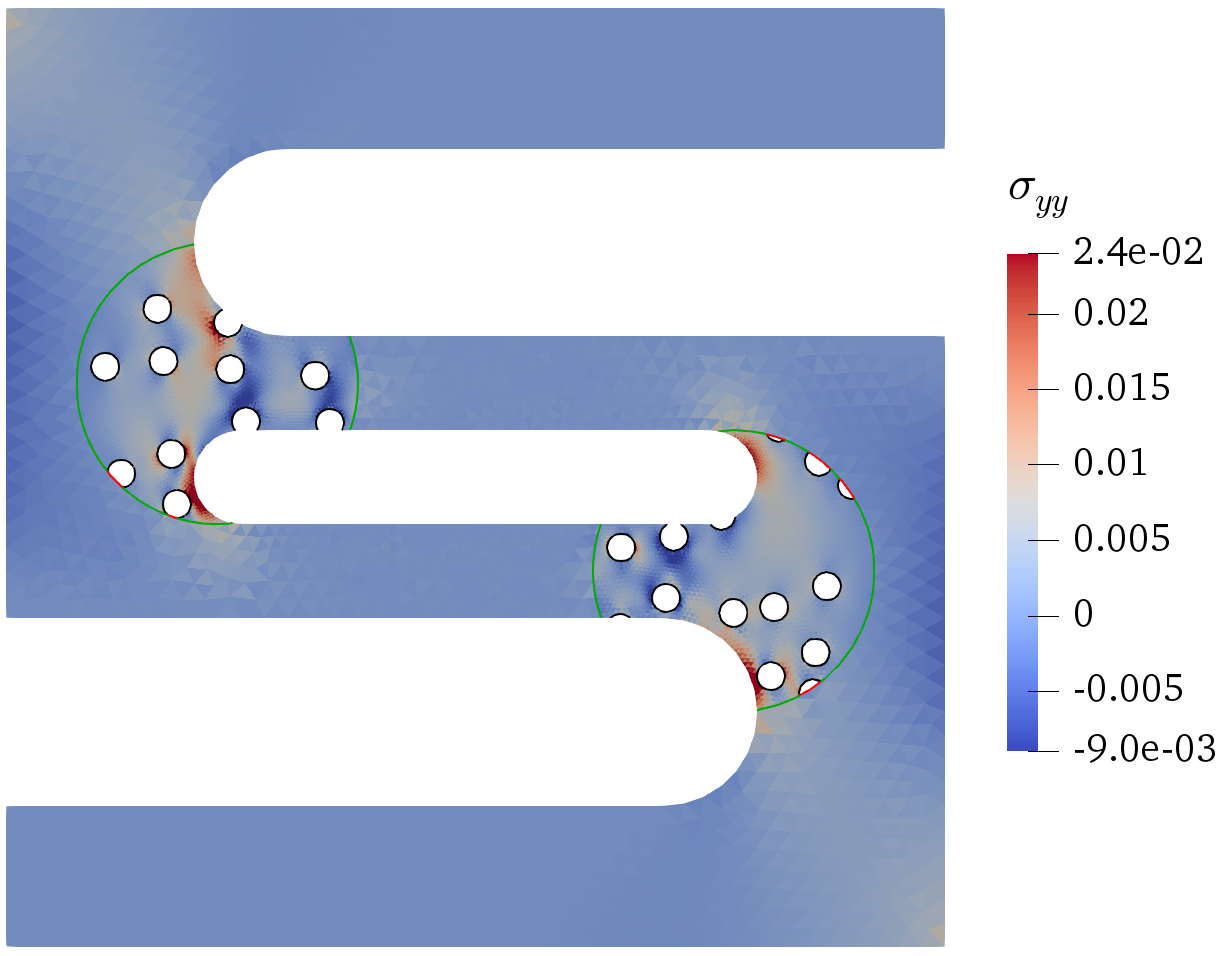}
     }
    \hfill
    \subfloat[ \label{SigMultiB}]{%
     \includegraphics[scale=0.32]{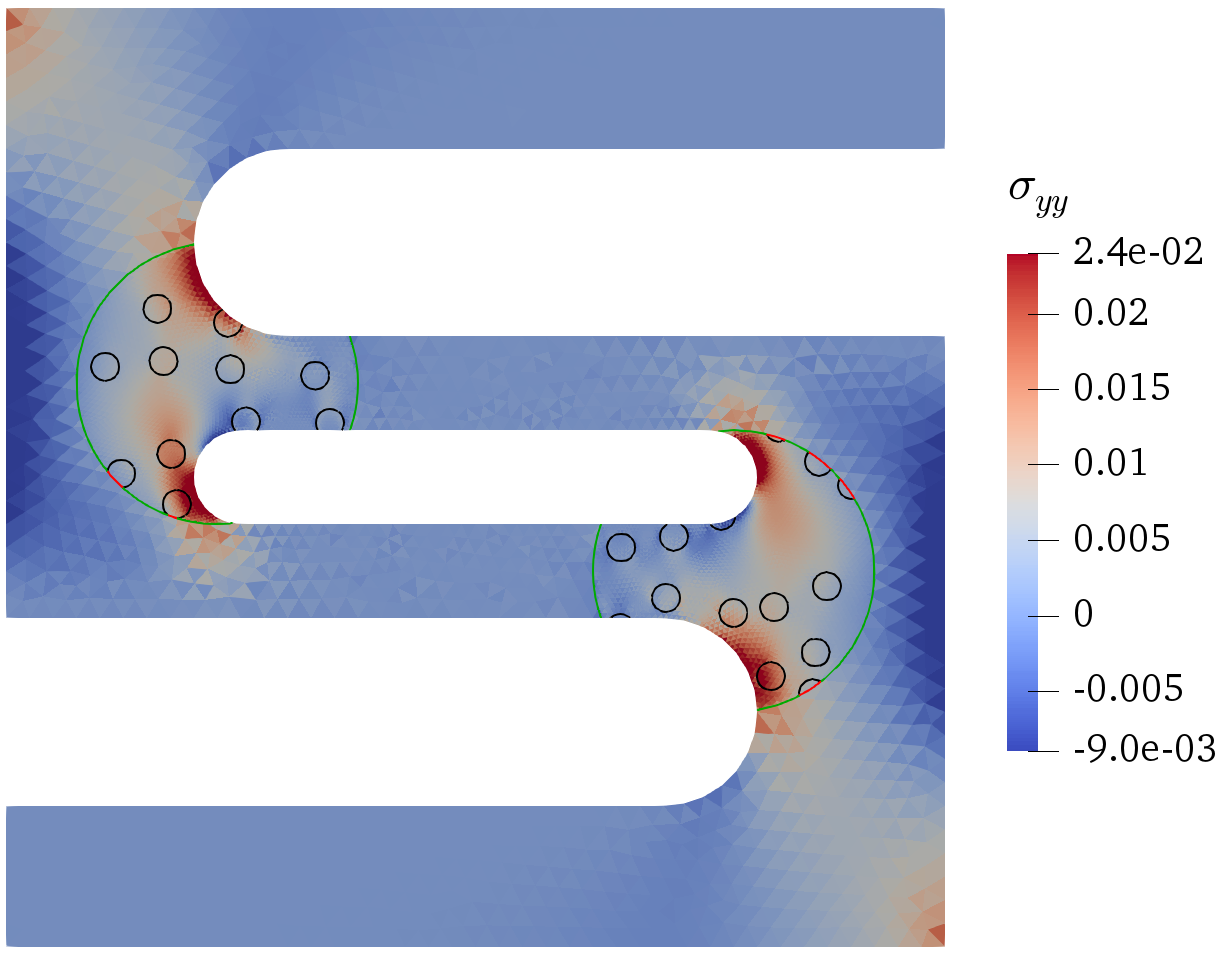}
     }
    \hfill
     \caption{Normal stress $\sigma_{yy}$ for the last time step; a) heterogeneities are voids, b) heterogeneities are hard inclusions}
     \label{fig:model2stress}
\end{figure}

Next, we solve the nonlinear problem and assess the corresponding solution fields.
The displacement field component $u_y$  for two models in Figure \ref{fig:model2disp} is smooth, especially in cut elements. Moreover, the stress field component $\sigma_{yy}$ shown in Figure \ref{fig:model2stress}, is smooth for both structures. However, as expected, the structure with hard inclusion inherits more stiffness and absorbs more stresses inside and outside the zoom.

\clearpage

\subsection{S shape porous structure with a time-dependent zoom}
\begin{figure}[h]
   \centering
     \subfloat[ \label{meshmov1}]{%
      \includegraphics[scale=0.3]{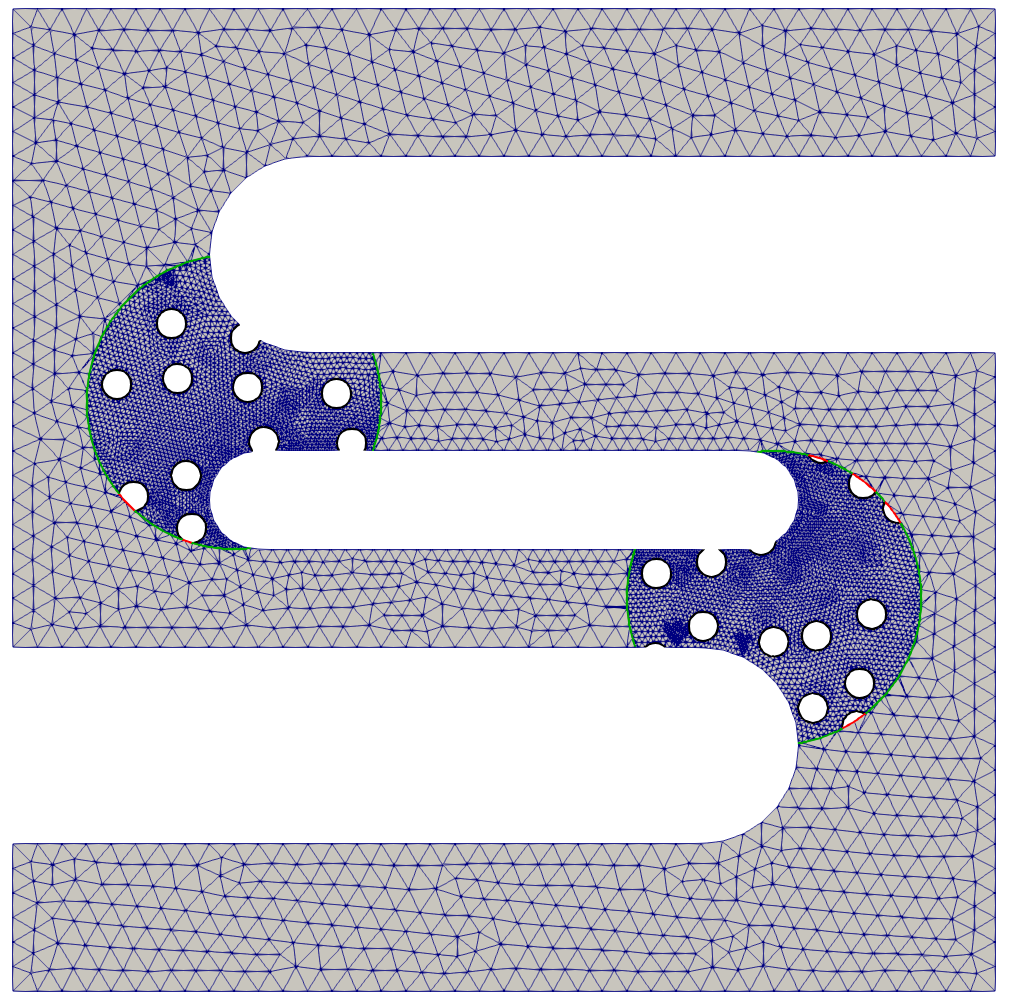}
     }
    \hfill
    \subfloat[ \label{meshmov2}]{%
     \includegraphics[scale=0.3]{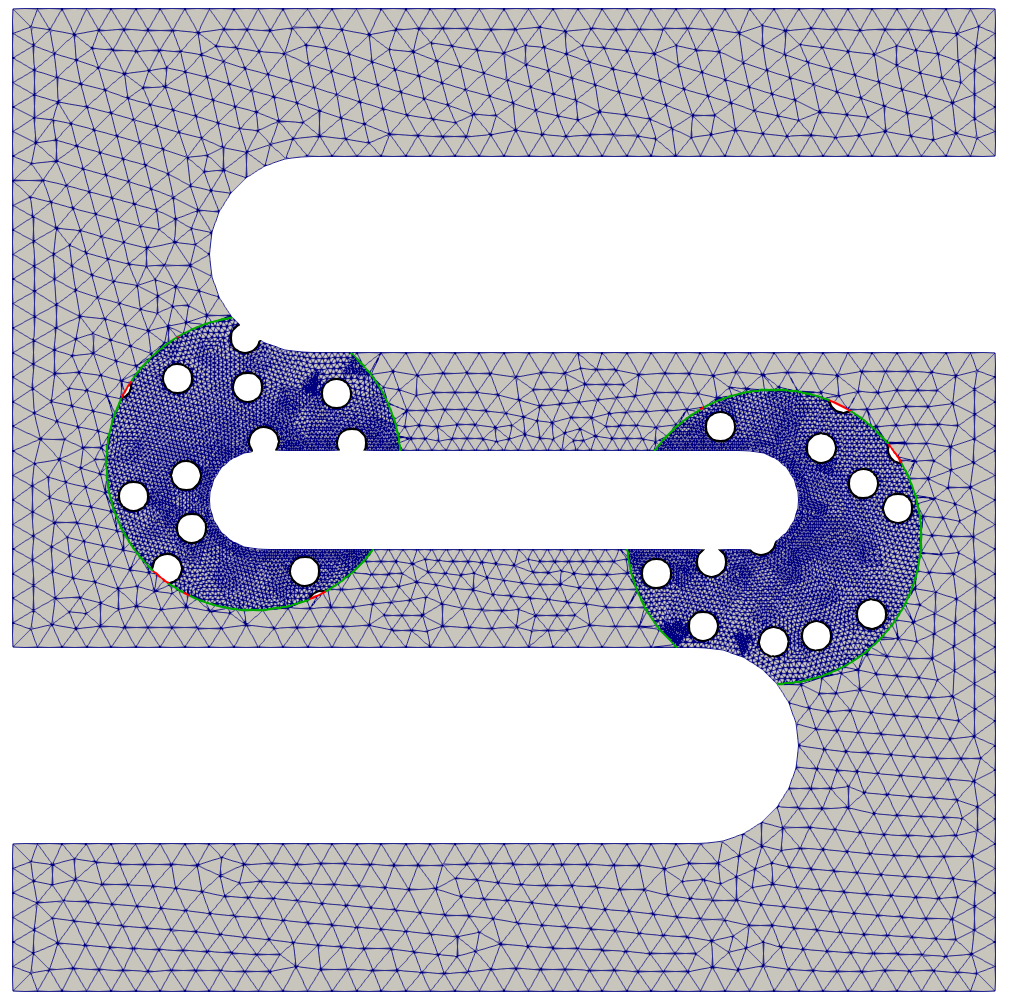}
     }
    \hfill
        \subfloat[ \label{meshmov3}]{%
     \includegraphics[scale=0.3]{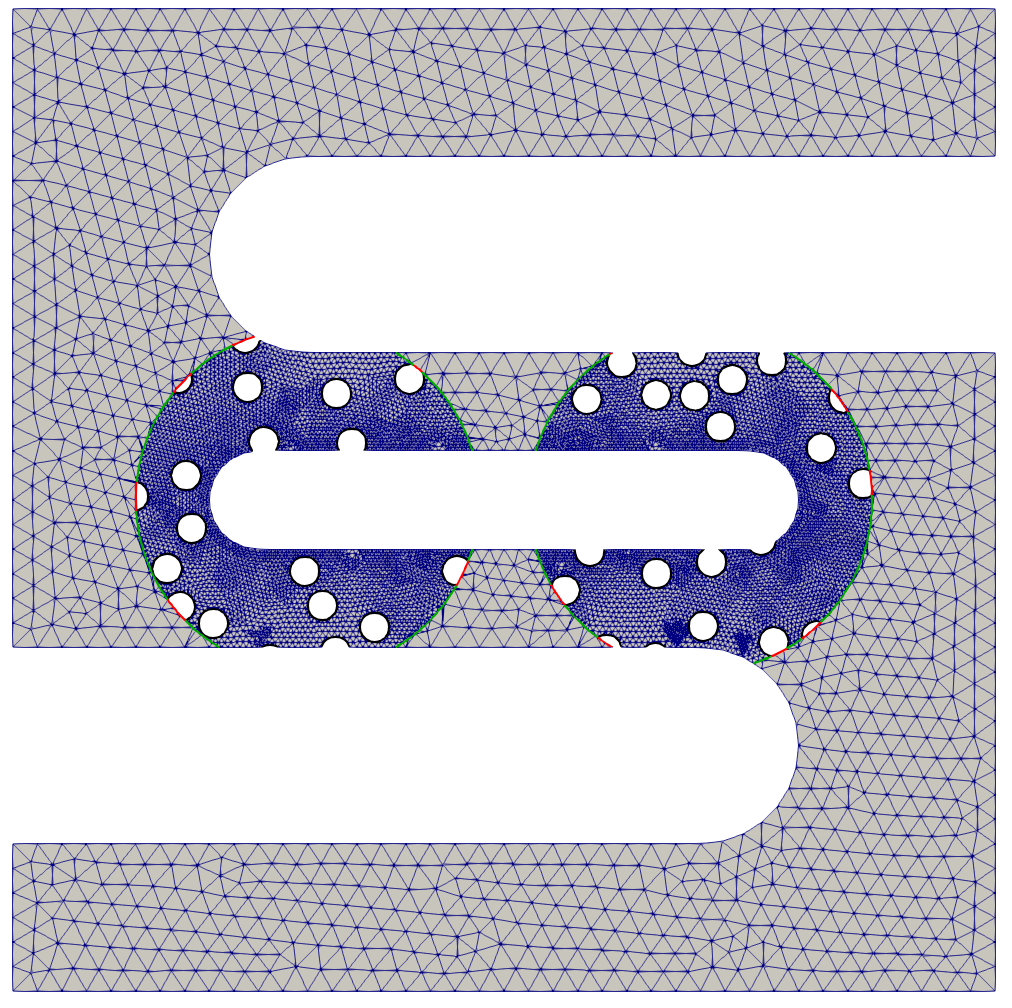}
     }
    \hfill
            \subfloat[ \label{meshmov3}]{%
     \includegraphics[scale=0.3]{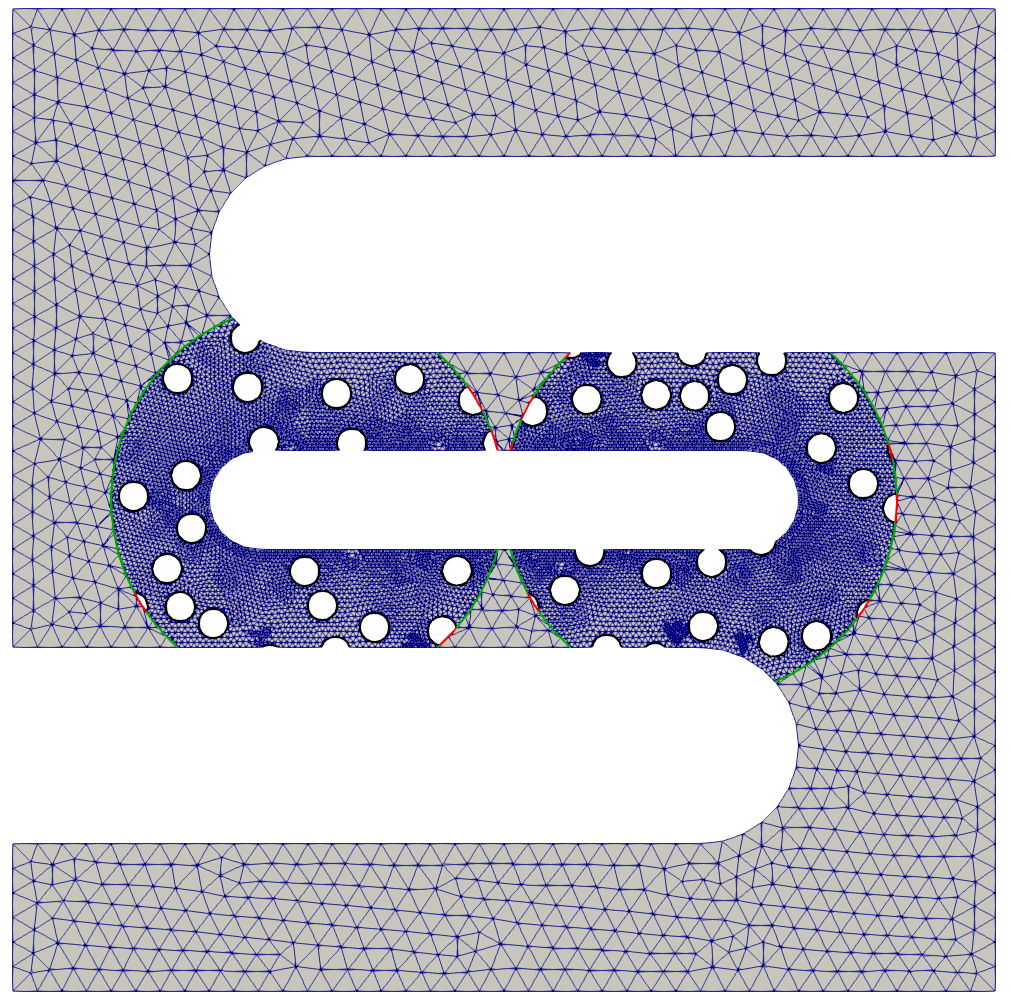}
     }
    \hfill
     \caption{Computational meshes for the microporous heterogeneous structure with different set of zooms at various time steps, a) $\tau = (0, 0.05)$, b) $\tau = (0, 0.1)$, c) $\tau = (0, 0.14)$ and d) $\tau = (0, 0.18)$. The cut elements here are depicted with their integration subtriangles.}
     \label{fig:model3mesh}
\end{figure}

This section is devoted to the numerical study of a time-dependent zooming approach for the proposed multiresolution framework with a von Mises plasticity material behaviour. We here consider the S shape microporous structure analysed in section \ref{Results:fixedSshape}. However, contrary to the previous section, we will not fix the zooms over the background mesh but relocate them during the simulation. As shown in Figure \ref{fig:model3mesh}, this relocation is carried out arbitrarily and independent of background mesh configuration. In this study, we change the location and size of zooming manually during the simulation to assess the numerical efficiency.

\begin{figure}[h]
   \centering
     \subfloat[ \label{dispmov1}]{%
      \includegraphics[scale=0.3]{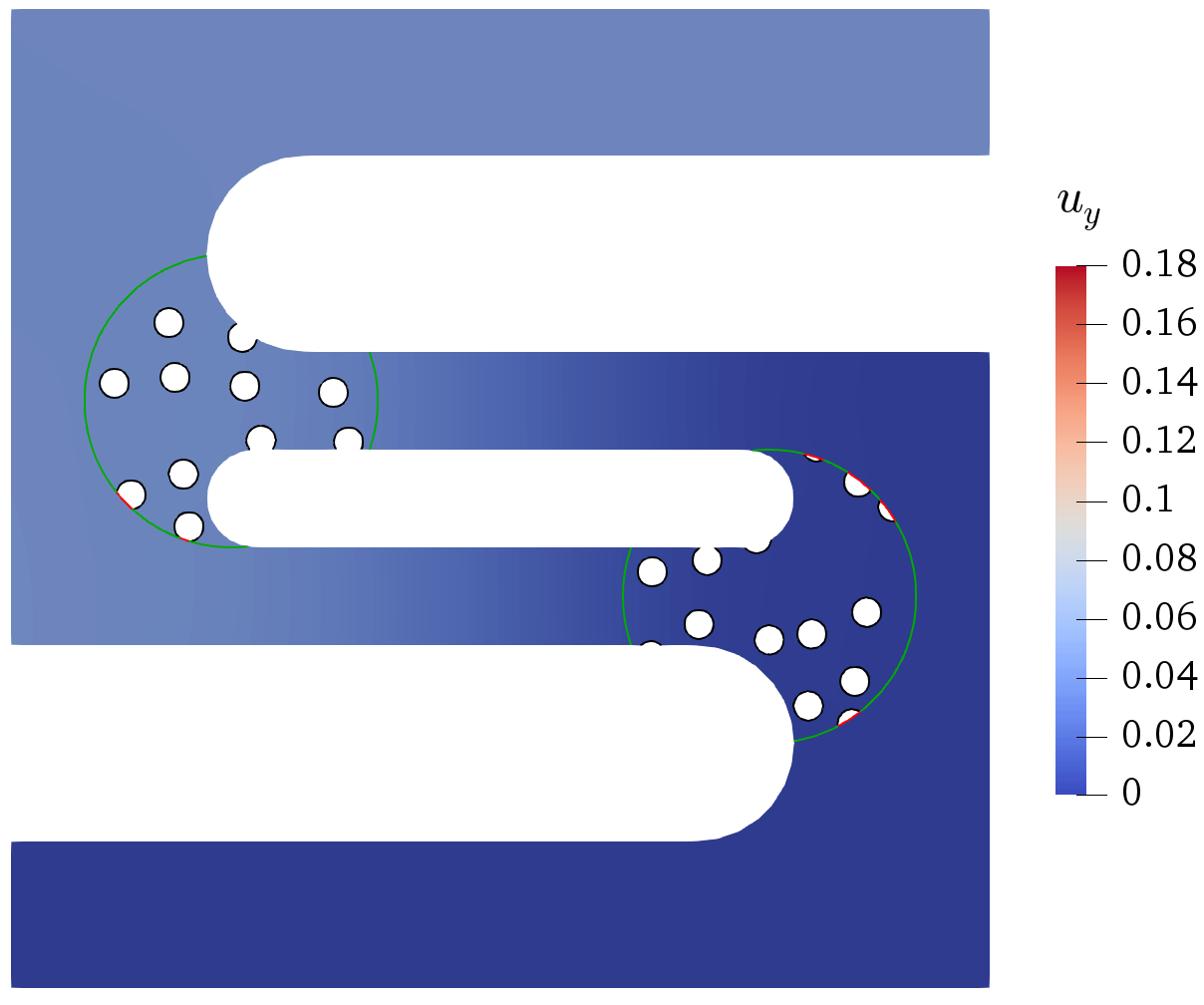}
     }
    \hfill
    \subfloat[ \label{dispmov2}]{%
     \includegraphics[scale=0.3]{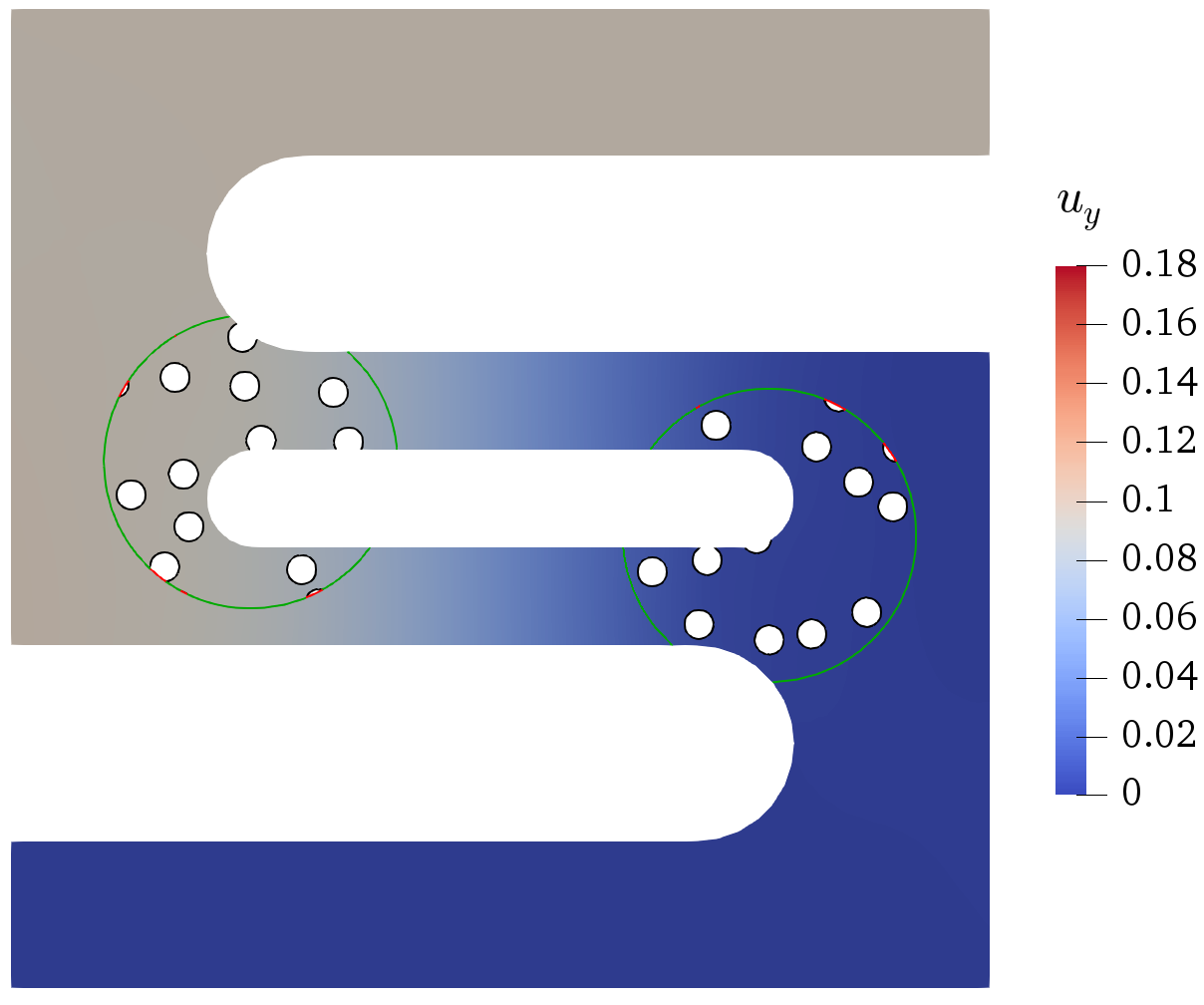}
     }
    \hfill
        \subfloat[ \label{dispmov3}]{%
     \includegraphics[scale=0.3]{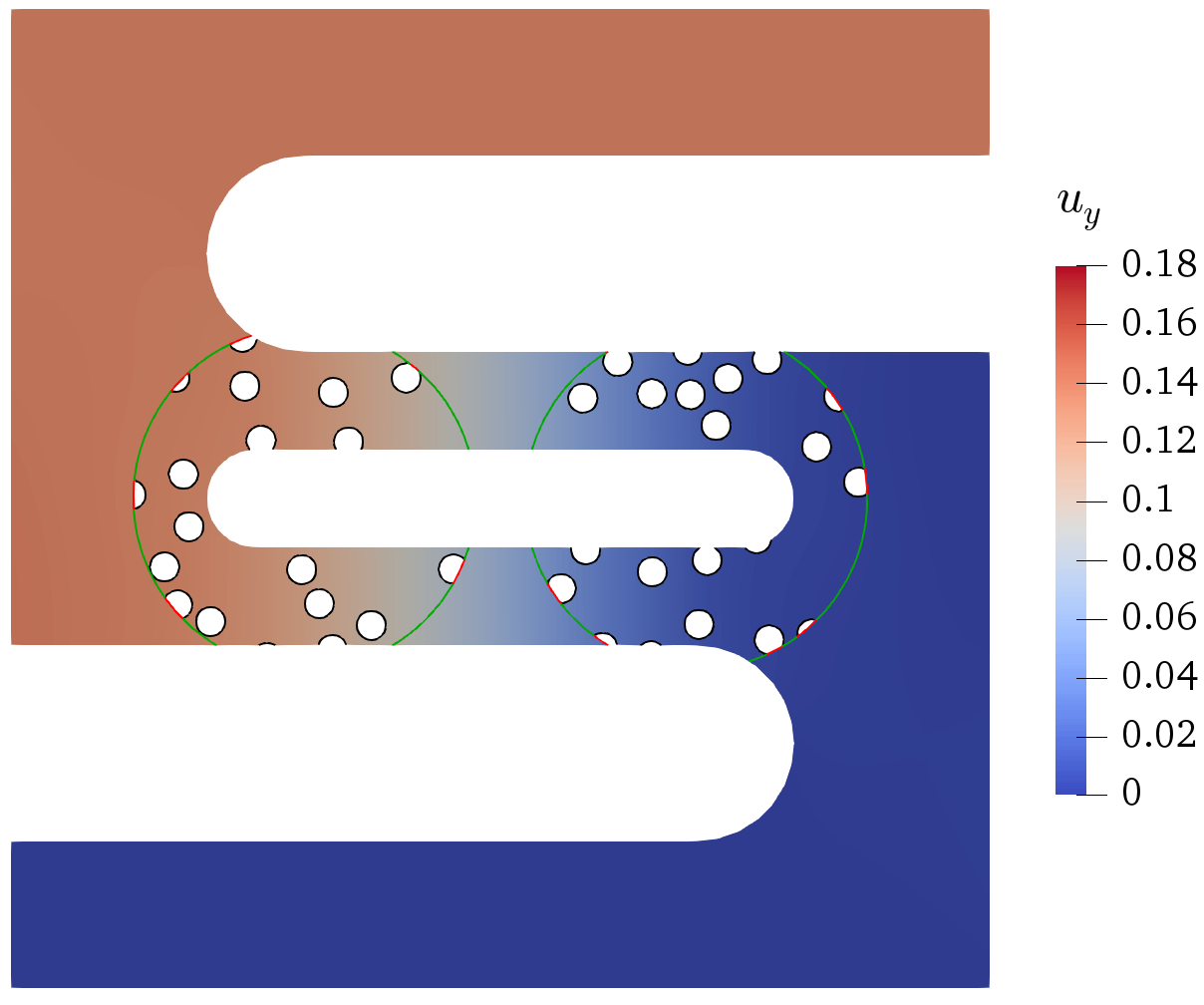}
     }
    \hfill
            \subfloat[ \label{dispmov3}]{%
     \includegraphics[scale=0.3]{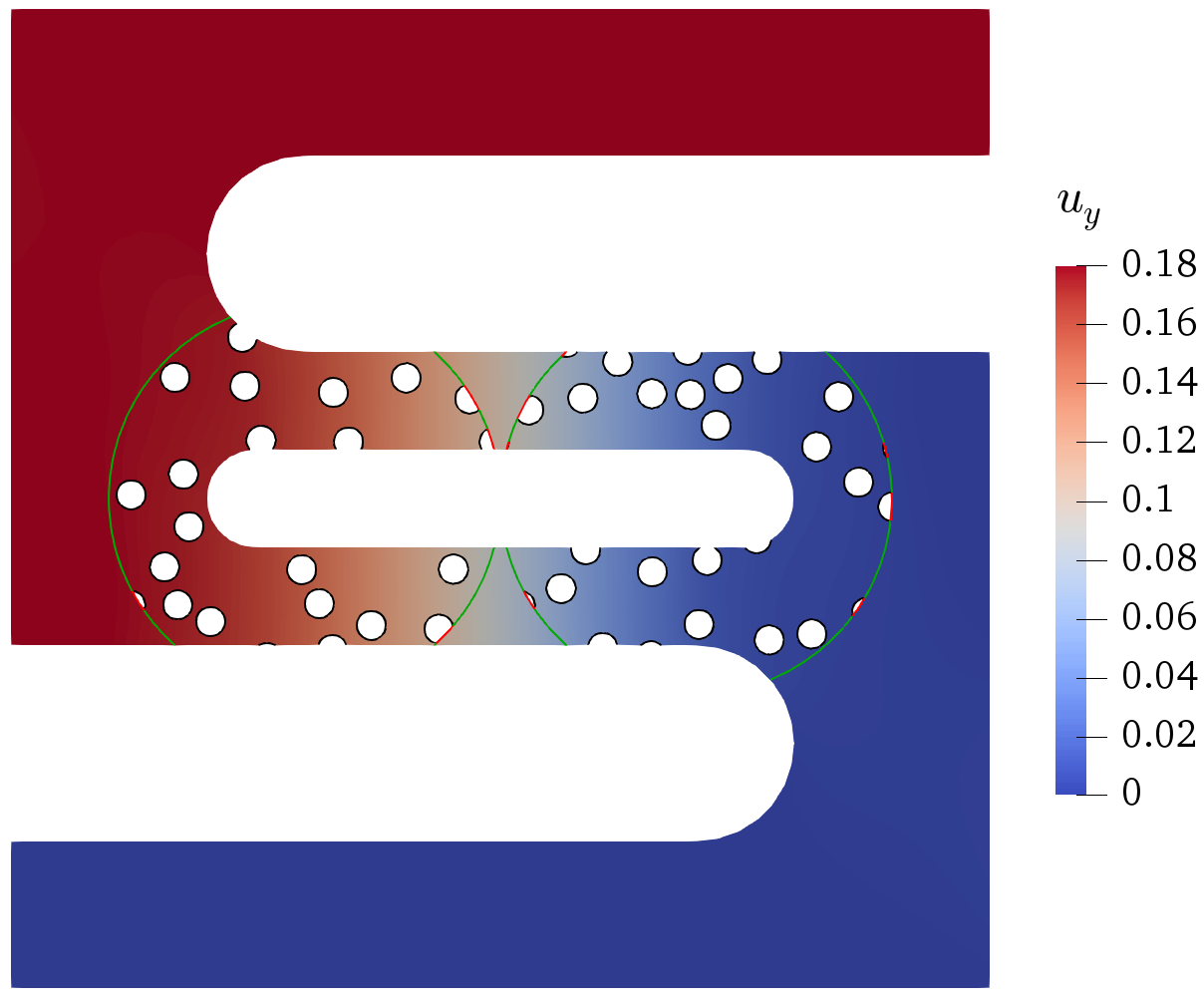}
     }
    \hfill
     \caption{Displacement component $u_y$ for the microporous heterogeneous structure with different set of zooms; a) $\tau = (0, 0.05)$, b) $\tau = (0, 0.1)$, c) $\tau = (0, 0.14)$ and d) $\tau = (0, 0.18)$.}
     \label{fig:model3disp}
\end{figure}

\begin{figure}[h]
   \centering
     \subfloat[ \label{sigmov1}]{%
      \includegraphics[scale=0.18]{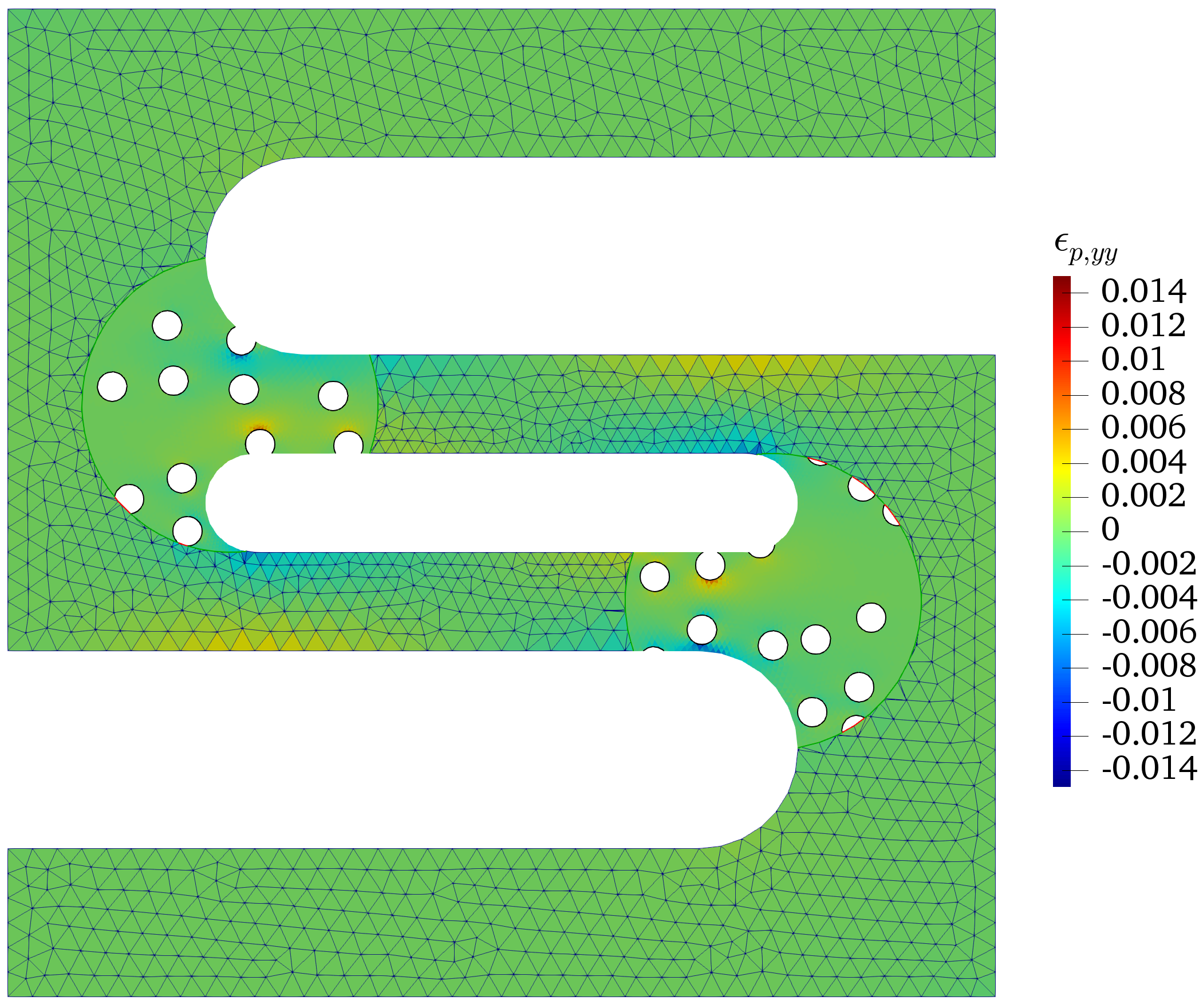}
     }
    \hfill
    \subfloat[ \label{sigmov2}]{%
     \includegraphics[scale=0.18]{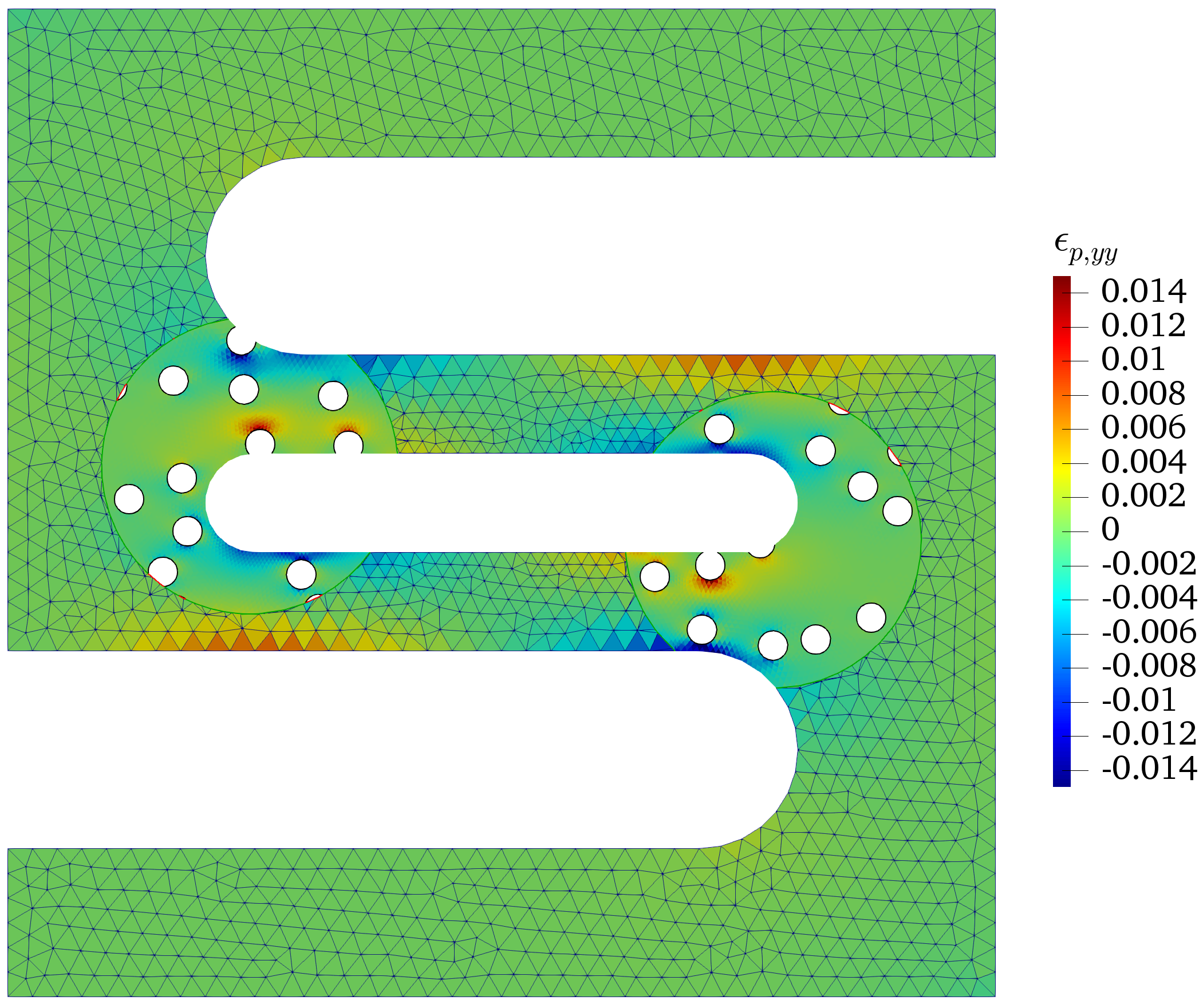}
     }
    \hfill
        \subfloat[ \label{sigmov3}]{%
     \includegraphics[scale=0.18]{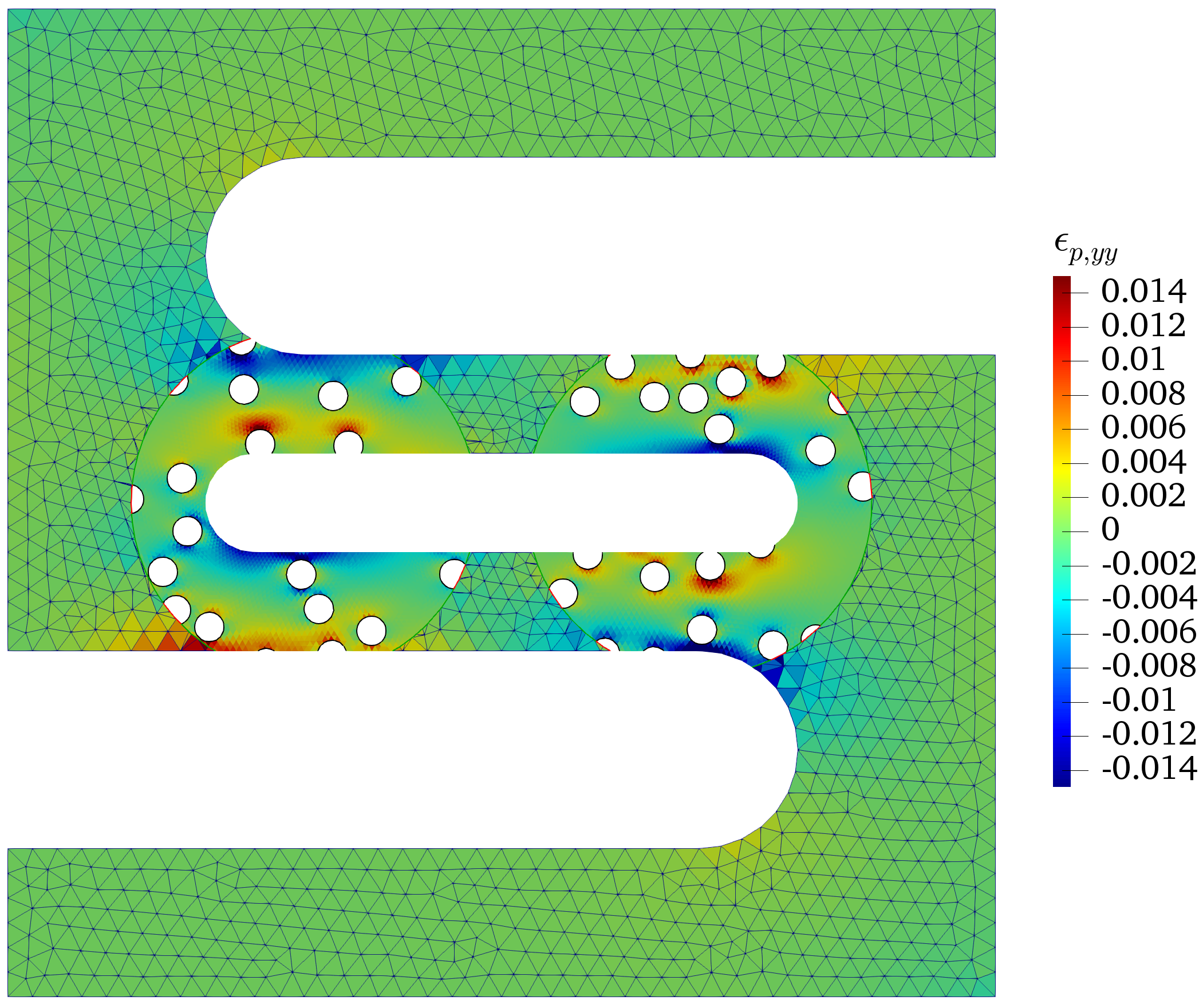}
     }
    \hfill
            \subfloat[ \label{sigmov3}]{%
     \includegraphics[scale=0.18]{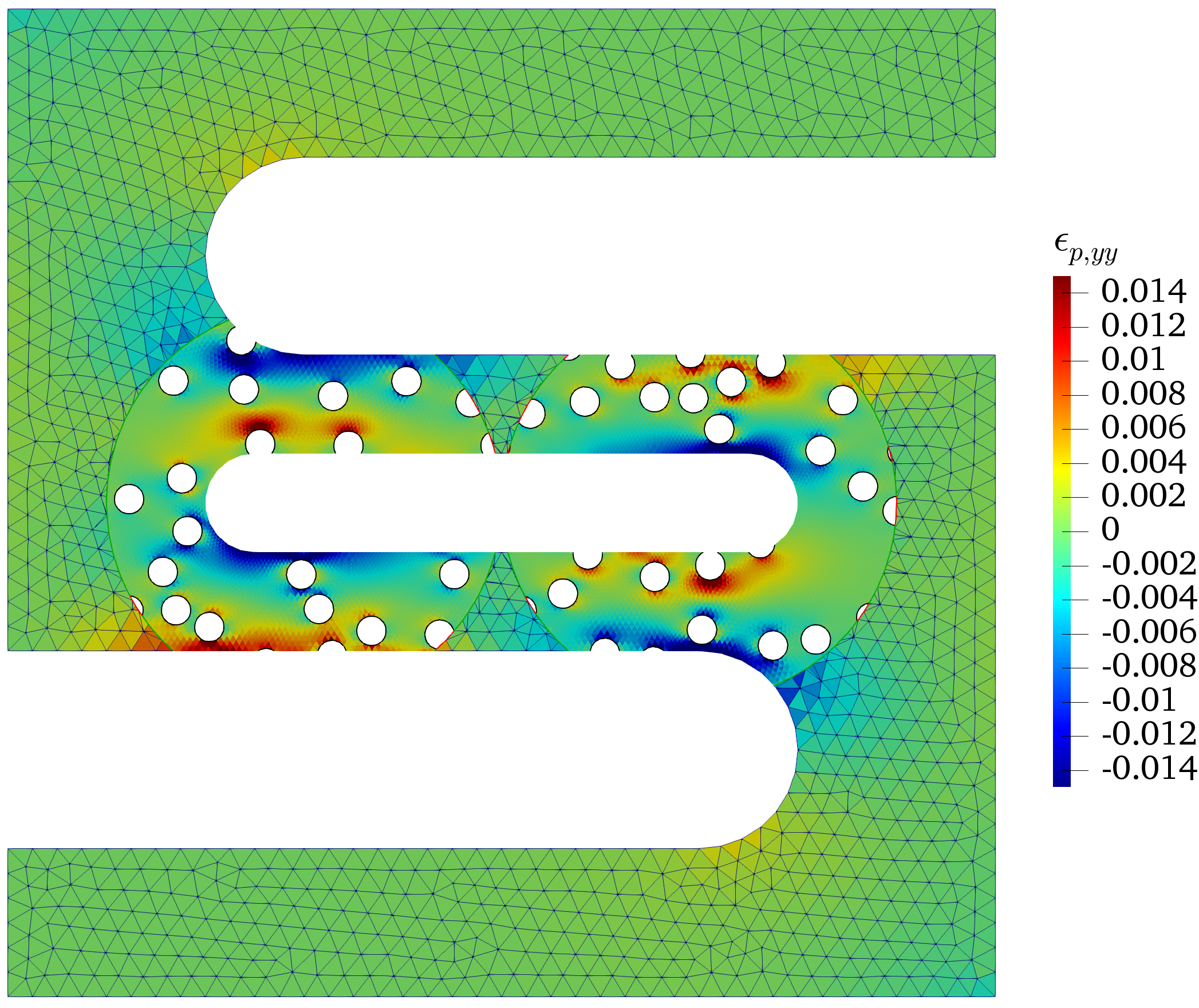}
     }
    \hfill
     \caption{Plastic strain component ${\epsilon}_{p,yy}$ contours at a) $\tau = (0, 0.05)$, b) $\tau = (0, 0.1)$, c) $\tau = (0, 0.14)$ and d) $\tau = (0, 0.18)$. Here, to ensure clarity in the presentation, we have chosen not to display the mesh inside the zoom regions. }
     \label{fig:model3sig}
\end{figure}

We present contour plots illustrating the displacement field component $u_y$ and the plastic strain component $\epsilon_{p,yy}$ at four distinct time instances, as depicted in Figures \ref{fig:model3disp} and \ref{fig:model3sig}, respectively. Our results demonstrate the stability of the multiscale solutions, devoid of any discernible oscillations during the relocation of the zoom region. The deliberate modification of the zoomed area throughout the simulation aims to accurately capture the growth of plastic deformations within the designated regions of interest.

However, to further enhance the accuracy of our approach in this regard, it is imperative to define the zoom level set function based on physics-based criteria using suitable error estimators. Incorporating such criteria into our methodology warrants future investigations to achieve improved precision. Thus, we propose that future works focus on exploring and implementing efficient error estimators to augment the fidelity of our results and provide more robust assessments of progressive phenomena (such as plasticity, damage and fracture) within the designated zoom regions.




\clearpage

\section{Conclusions}

In this study, we have presented a robust concurrent multiscale framework for modelling heterogeneous structures in a mesh-independent manner, utilising the CutFEM algorithm. The framework considers complex geometries of microscale model (inside the zoom region) and macroscale model (outside of the zoom region) seamlessly by proposing a versatile multiple level set approach, which defines three types of arbitrary interfaces over a fixed background mesh; the interface representing the microstructure architecture, the interface between the microscale matrix and the macroscale, and the interface between the micro-pores (or micro-inclusions) and the macroscale. To preserve accuracy inside the zoom, the background mesh is hierarchically refined in this region, which leads to an unfitted multiresolution framework adopted with a concurrent multiscale solver.

To capture the discontinuities in the solution fields across the interfaces, we have utilised the CutFEM enrichment technique, while implementing ghost penalty regularisation terms to prevent the ill-conditioning of the multiscale system matrix, particularly when cuts are near the nodes. Furthermore, we have used Nitsche's technique to glue together the subdomains.

To validate our methodology, we compared our multiscale model results with a full microscale FEM, which demonstrated that our unfitted concurrent multiscale framework yields accurate results, particularly inside the zooming region. In addition, we have demonstrated the high geometrical flexibility of our framework by testing two different types of microstructures, which were comprised of either micro-pores or micro-inclusions, and by considering both linear elastic and nonlinear plasticity constitutive models. The results have shown that our framework is capable of representing various types of arbitrary interfaces and performing corresponding simultaneous enrichments successfully.

Our contribution provides a strong foundation for further developments in the field of computational mechanics. However, we acknowledge that our assumption of the zooming region being known in advance may not always be feasible, particularly when modelling progressive phenomena such as crack propagation in heterogeneous materials. Therefore, future work may involve extending our methodology to handle scenarios where the zooming region is not known in advance.

\section*{Acknowledgement}

The authors acknowledge the support of Cardiff University on this work, which was funded by the European Union’s Horizon 2020 research and innovation program under the Marie Sklodowska-Curie grant agreement No. 764644.


\bibliography{HierarchicalRefs}

\begin{thebibliography}{10}
\expandafter\ifx\csname url\endcsname\relax
  \def\url#1{\texttt{#1}}\fi
\expandafter\ifx\csname urlprefix\endcsname\relax\def\urlprefix{URL }\fi
\expandafter\ifx\csname href\endcsname\relax
  \def\href#1#2{#2} \def\path#1{#1}\fi

\bibitem{tadmor_miller_2011}
E.~B. Tadmor, R.~E. Miller, Modeling Materials: Continuum, Atomistic and
  Multiscale Techniques, Cambridge University Press, 2011.
\newblock \href {https://doi.org/10.1017/CBO9781139003582}
  {\path{doi:10.1017/CBO9781139003582}}.

\bibitem{Kanoute.09}
P.~Kanoute, D.~Boso, J.-L. Chaboche, B.~Schrefler, Multiscale methods for
  composites: A review, Archives of Computational Methods in Engineering 16
  (2009) 31--75.
\newblock \href {https://doi.org/https://doi.org/10.1007/s11831-008-9028-8}
  {\path{doi:https://doi.org/10.1007/s11831-008-9028-8}}.

\bibitem{FEYEL03}
F.~Feyel, A multilevel finite element method (fe2) to describe the response of
  highly non-linear structures using generalized continua, Computer Methods in
  Applied Mechanics and Engineering 192~(28) (2003) 3233--3244, multiscale
  Computational Mechanics for Materials and Structures.
\newblock \href {https://doi.org/https://doi.org/10.1016/S0045-7825(03)00348-7}
  {\path{doi:https://doi.org/10.1016/S0045-7825(03)00348-7}}.

\bibitem{tallec1994domain}
P.~L. Tallec, \href{https://cir.nii.ac.jp/crid/1571980074848734464}{Domain
  decomposition methods in computational mechnaics}, Computational Mechnaics
  Advances 1 (1994) 121--220.
\newline\urlprefix\url{https://cir.nii.ac.jp/crid/1571980074848734464}

\bibitem{toselli2004domain}
A.~Toselli, O.~Widlund, Domain decomposition methods-algorithms and theory,
  Vol.~34, Springer Science \& Business Media, 2004.

\bibitem{zohdi1996hierarchical}
T.~I. Zohdi, J.~T. Oden, G.~J. Rodin, Hierarchical modeling of heterogeneous
  bodies, Computer methods in applied mechanics and engineering 138~(1-4)
  (1996) 273--298.
\newblock \href {https://doi.org/https://doi.org/10.1016/S0045-7825(96)01106-1}
  {\path{doi:https://doi.org/10.1016/S0045-7825(96)01106-1}}.

\bibitem{ZOHDI19992507}
T.~Zohdi, P.~Wriggers, A domain decomposition method for bodies with
  heterogeneous microstructure basedon material regularization, International
  Journal of Solids and Structures 36~(17) (1999) 2507--2525.
\newblock \href {https://doi.org/https://doi.org/10.1016/S0020-7683(98)00124-3}
  {\path{doi:https://doi.org/10.1016/S0020-7683(98)00124-3}}.

\bibitem{ZOHDI2001}
T.~Zohdi, P.~Wriggers, C.~Huet, A method of substructuring large-scale
  computational micromechanical problems, Computer Methods in Applied Mechanics
  and Engineering 190~(43) (2001) 5639--5656.
\newblock \href {https://doi.org/https://doi.org/10.1016/S0045-7825(01)00189-X}
  {\path{doi:https://doi.org/10.1016/S0045-7825(01)00189-X}}.

\bibitem{RAGHAVAN2004497}
P.~Raghavan, S.~Ghosh, Concurrent multi-scale analysis of elastic composites by
  a multi-level computational model, Computer Methods in Applied Mechanics and
  Engineering 193~(6) (2004) 497--538.
\newblock \href {https://doi.org/https://doi.org/10.1016/j.cma.2003.10.007}
  {\path{doi:https://doi.org/10.1016/j.cma.2003.10.007}}.

\bibitem{Dhia.Rateau.05}
H.~Dhia, G.~Rateau, The arlequin method as a flexible engineering design tool,
  International Journal for Numerical Methods in Engineering 62 (2005)
  1442--1462.
\newblock \href {https://doi.org/https://doi.org/10.1002/nme.1229}
  {\path{doi:https://doi.org/10.1002/nme.1229}}.

\bibitem{BURMAN20102680}
E.~Burman, P.~Hansbo, Fictitious domain finite element methods using cut
  elements: I. a stabilized lagrange multiplier method, Computer Methods in
  Applied Mechanics and Engineering 199~(41) (2010) 2680 -- 2686.
\newblock \href {https://doi.org/https://doi.org/10.1016/j.cma.2010.05.011}
  {\path{doi:https://doi.org/10.1016/j.cma.2010.05.011}}.

\bibitem{Ji04}
H.~Ji, J.~Dolbow, On strategies for enforcing interfacial constraints and
  evaluating jump conditions with the extended finite element method,
  International Journal for Numerical Methods in Engineering 61~(14) (2004)
  2508--2535.
\newblock \href {https://doi.org/https://doi.org/10.1002/nme.1167}
  {\path{doi:https://doi.org/10.1002/nme.1167}}.

\bibitem{dsouza21}
S.~M. Dsouza, A.~Pramod, E.~T. Ooi, C.~Song, S.~Natarajan, Robust modelling of
  implicit interfaces by the scaled boundary finite element method, Engineering
  Analysis with Boundary Elements 124 (2021) 266--286.
\newblock \href
  {https://doi.org/https://doi.org/10.1016/j.enganabound.2020.12.025}
  {\path{doi:https://doi.org/10.1016/j.enganabound.2020.12.025}}.

\bibitem{BURMAN2012328}
E.~Burman, P.~Hansbo,
  \href{http://www.sciencedirect.com/science/article/pii/S0168927411000298}{Fictitious
  domain finite element methods using cut elements: Ii. a stabilized nitsche
  method}, Applied Numerical Mathematics 62~(4) (2012) 328 -- 341, third
  Chilean Workshop on Numerical Analysis of Partial Differential Equations
  (WONAPDE 2010).
\newblock \href {https://doi.org/https://doi.org/10.1016/j.apnum.2011.01.008}
  {\path{doi:https://doi.org/10.1016/j.apnum.2011.01.008}}.
\newline\urlprefix\url{http://www.sciencedirect.com/science/article/pii/S0168927411000298}

\bibitem{CAI2021113880}
Y.~Cai, J.~Chen, N.~Wang, A nitsche extended finite element method for the
  biharmonic interface problem, Computer Methods in Applied Mechanics and
  Engineering 382 (2021) 113880.
\newblock \href {https://doi.org/https://doi.org/10.1016/j.cma.2021.113880}
  {\path{doi:https://doi.org/10.1016/j.cma.2021.113880}}.

\bibitem{liu2014non}
Y.~Liu, Q.~Sun, X.~Fan, A non-intrusive global/local algorithm with
  non-matching interface: derivation and numerical validation, Computer Methods
  in Applied Mechanics and Engineering 277 (2014) 81--103.
\newblock \href {https://doi.org/https://doi.org/10.1016/j.cma.2014.04.012}
  {\path{doi:https://doi.org/10.1016/j.cma.2014.04.012}}.

\bibitem{Akbari15}
A.~Akbari-Rahimabadi, P.~Kerfriden, S.~Bordas, Scale selection in nonlinear
  fracture mechanics of heterogeneous materials, Philosophical Magazine 95
  (2015) 3328--3347.
\newblock \href {https://doi.org/https://doi.org/10.1080/14786435.2015.1061716}
  {\path{doi:https://doi.org/10.1080/14786435.2015.1061716}}.

\bibitem{Xiao2004}
S.~Xiao, T.~Belytschko, A bridging domain method for coupling continua with
  molecular dynamics, Computer Methods in Applied Mechanics and Engineering 193
  (2004) 1645--1669.
\newblock \href {https://doi.org/https://doi.org/10.1016/j.cma.2003.12.053}
  {\path{doi:https://doi.org/10.1016/j.cma.2003.12.053}}.

\bibitem{Tadmor1996}
E.~B. Tadmor, M.~Ortiz, R.~Phillips, Quasicontinuum analysis of defects in
  solids, Philosophical Magazine A 73~(6) (1996) 1529--1563.
\newblock \href {https://doi.org/https://doi.org/10.1080/01418619608243000}
  {\path{doi:https://doi.org/10.1080/01418619608243000}}.

\bibitem{Beex.Kerfriden.ea.14}
L.~Beex, P.~Kerfriden, T.~Rabczuk, S.~Bordas, Quasicontinuum-based multiscale
  approaches for plate-like beam lattices experiencing in-plane and
  out-of-plane deformation, Computer Methods in Applied Mechanics and
  Engineering 279 (2014) 348–378.
\newblock \href {https://doi.org/https://doi.org/10.1016/j.cma.2014.06.018}
  {\path{doi:https://doi.org/10.1016/j.cma.2014.06.018}}.

\bibitem{Dhia.98}
H.~B. Dhia, Multiscale mechanical problems: the arlequin method, Comptes Rendus
  de l'Academie des Sciences Series IIB Mechanics Physics Astronomy 326 (1998)
  899--904.

\bibitem{Lamichhane.Wohlmuth.04}
B.~P. Lamichhane, B.~I. Wohlmuth, Mortar finite element for interface problems,
  Computing 72 (2004) 333--348.
\newblock \href {https://doi.org/https://doi.org/10.1007/s00607-003-0062-y}
  {\path{doi:https://doi.org/10.1007/s00607-003-0062-y}}.

\bibitem{Burman.Claus.ea.15}
E.~Burman, S.~Claus, P.~Hansbo, M.~Larson, A.~Massing, Cutfem: discretizing
  geometry and partial differential equations, International Journal for
  Numerical Methods in Engineering 104 (2015) 472--501.
\newblock \href {https://doi.org/https://doi.org/10.1002/nme.4823}
  {\path{doi:https://doi.org/10.1002/nme.4823}}.

\bibitem{Parvizian07}
J.~Parvizian, A.~Düster, E.~Rank, Finite cell method, Computational Mechanics
  41 (2007) 121–133.
\newblock \href {https://doi.org/https://doi.org/10.1007/s00466-007-0173-y}
  {\path{doi:https://doi.org/10.1007/s00466-007-0173-y}}.

\bibitem{duster2008finite}
A.~D{\"u}ster, J.~Parvizian, Z.~Yang, E.~Rank, The finite cell method for
  three-dimensional problems of solid mechanics, Computer methods in applied
  mechanics and engineering 197~(45-48) (2008) 3768--3782.
\newblock \href {https://doi.org/https://doi.org/10.1016/j.cma.2008.02.036}
  {\path{doi:https://doi.org/10.1016/j.cma.2008.02.036}}.

\bibitem{nadal2013efficient}
E.~Nadal, J.~R{\'o}denas, J.~Albelda, M.~Tur, J.~Taranc{\'o}n, F.~Fuenmayor,
  Efficient finite element methodology based on cartesian grids: application to
  structural shape optimization, in: Abstract and applied analysis, Vol. 2013,
  Hindawi, 2013.
\newblock \href {https://doi.org/https://doi.org/10.1155/2013/953786}
  {\path{doi:https://doi.org/10.1155/2013/953786}}.

\bibitem{bernardi1989new}
C.~Bernardi, \href{https://cir.nii.ac.jp/crid/1570854175308447616}{A new
  nonconforming approach to domain decomposition: the mortar element method},
  Nonlinear partial equations and their applications (1989).
\newline\urlprefix\url{https://cir.nii.ac.jp/crid/1570854175308447616}

\bibitem{arbogast2000mixed}
T.~Arbogast, L.~C. Cowsar, M.~F. Wheeler, I.~Yotov, Mixed finite element
  methods on nonmatching multiblock grids, SIAM Journal on Numerical Analysis
  37~(4) (2000) 1295--1315.
\newblock \href {https://doi.org/https://doi.org/10.1137/S0036142996308447}
  {\path{doi:https://doi.org/10.1137/S0036142996308447}}.

\bibitem{wohlmuth2000mortar}
B.~I. Wohlmuth, A mortar finite element method using dual spaces for the
  lagrange multiplier, SIAM journal on numerical analysis 38~(3) (2000)
  989--1012.
\newblock \href {https://doi.org/https://doi.org/10.1137/S0036142999350929}
  {\path{doi:https://doi.org/10.1137/S0036142999350929}}.

\bibitem{park2002simple}
K.~Park, C.~Felippa, G.~Rebel, A simple algorithm for localized construction of
  non-matching structural interfaces, International Journal for Numerical
  Methods in Engineering 53~(9) (2002) 2117--2142.
\newblock \href {https://doi.org/https://doi.org/10.1002/nme.374}
  {\path{doi:https://doi.org/10.1002/nme.374}}.

\bibitem{arbogast2007multiscale}
T.~Arbogast, G.~Pencheva, M.~F. Wheeler, I.~Yotov, A multiscale mortar mixed
  finite element method, Multiscale Modeling \& Simulation 6~(1) (2007)
  319--346.
\newblock \href {https://doi.org/https://doi.org/10.1137/060662587}
  {\path{doi:https://doi.org/10.1137/060662587}}.

\bibitem{subber2016asynchronous}
W.~Subber, K.~Matou{\v{s}}, Asynchronous space--time algorithm based on a
  domain decomposition method for structural dynamics problems on non-matching
  meshes, Computational Mechanics 57 (2016) 211--235.
\newblock \href {https://doi.org/https://doi.org/10.1007/s00466-015-1228-0}
  {\path{doi:https://doi.org/10.1007/s00466-015-1228-0}}.

\bibitem{Moes99}
N.~Moës, J.~Dolbow, T.~Belytschko, A finite element method for crack growth
  without remeshing, International Journal for Numerical Methods in Engineering
  46~(1) (1999) 131--150.
\newblock \href
  {https://doi.org/https://doi.org/10.1002/(SICI)1097-0207(19990910)46:1<131::AID-NME726>3.0.CO;2-J}
  {\path{doi:https://doi.org/10.1002/(SICI)1097-0207(19990910)46:1<131::AID-NME726>3.0.CO;2-J}}.

\bibitem{Burman.10}
E.~Burman, Ghost penalty, Comptes Rendus Mathematique 348 (2010) 1217--1220.
\newblock \href {https://doi.org/https://doi.org/10.1016/j.crma.2010.10.006}
  {\path{doi:https://doi.org/10.1016/j.crma.2010.10.006}}.

\bibitem{hansbo2002unfitted}
A.~Hansbo, P.~Hansbo, An unfitted finite element method, based on nitsche’s
  method, for elliptic interface problems, Computer methods in applied
  mechanics and engineering 191~(47-48) (2002) 5537--5552.
\newblock \href {https://doi.org/https://doi.org/10.1016/S0045-7825(02)00524-8}
  {\path{doi:https://doi.org/10.1016/S0045-7825(02)00524-8}}.

\bibitem{becker2009nitsche}
R.~Becker, E.~Burman, P.~Hansbo, A nitsche extended finite element method for
  incompressible elasticity with discontinuous modulus of elasticity, Computer
  Methods in Applied Mechanics and Engineering 198~(41-44) (2009) 3352--3360.
\newblock \href {https://doi.org/https://doi.org/10.1016/j.cma.2009.06.017}
  {\path{doi:https://doi.org/10.1016/j.cma.2009.06.017}}.

\bibitem{talebi2013molecular}
H.~Talebi, M.~Silani, S.~Bordas, P.~Kerfriden, T.~Rabczuk, Molecular
  dynamics/xfem coupling by a three-dimensional extended bridging domain with
  applications to dynamic brittle fracture, International Journal for
  Multiscale Computational Engineering 11 (2013) 527--541.
\newblock \href {https://doi.org/10.1615/IntJMultCompEng.2013005838}
  {\path{doi:10.1615/IntJMultCompEng.2013005838}}.

\bibitem{Mikaeili18}
E.~Mikaeili, B.~Schrefler, Xfem, strong discontinuities and second-order work
  in shear band modeling of saturated porous media, Acta Geotechnica 13 (2018)
  1249--1264.
\newblock \href {https://doi.org/https://doi.org/10.1007/s11440-018-0734-6}
  {\path{doi:https://doi.org/10.1007/s11440-018-0734-6}}.

\bibitem{bordas2023partition}
S.~Bordas, A.~Menk,
  \href{https://books.google.co.uk/books?id=2ygSywAACAAJ}{Partition of Unity
  Methods}, Wiley, 2023.
\newline\urlprefix\url{https://books.google.co.uk/books?id=2ygSywAACAAJ}

\bibitem{Claus19.2}
S.~Claus, P.~Kerfriden, A cutfem method for two-phase flow problems, Computer
  Methods in Applied Mechanics and Engineering 348 (2019) 185--206.
\newblock \href {https://doi.org/10.1016/j.cma.2019.01.009}
  {\path{doi:10.1016/j.cma.2019.01.009}}.

\bibitem{FRACHON201977}
A cut finite element method for incompressible two-phase navier–stokes flows,
  Journal of Computational Physics 384 (2019) 77--98.
\newblock \href {https://doi.org/https://doi.org/10.1016/j.jcp.2019.01.028}
  {\path{doi:https://doi.org/10.1016/j.jcp.2019.01.028}}.

\bibitem{hansbo2016cut}
P.~Hansbo, M.~G. Larson, S.~Zahedi, A cut finite element method for coupled
  bulk-surface problems on time-dependent domains, Computer Methods in Applied
  Mechanics and Engineering 307 (2016) 96--116.
\newblock \href {https://doi.org/https://doi.org/10.1016/j.cma.2016.04.012}
  {\path{doi:https://doi.org/10.1016/j.cma.2016.04.012}}.

\bibitem{Claus.Bigot.ea.18}
S.~Claus, S.~Bigot, P.~Kerfriden, Cutfem method for stefan-signorini problems
  with application in pulsed laser ablation, SIAM Journal on Scientific
  Computing 40 (2018) 1444--1469.
\newblock \href {https://doi.org/https://doi.org/10.1137/18M1185697}
  {\path{doi:https://doi.org/10.1137/18M1185697}}.

\bibitem{Claus.Kerfriden.18}
S.~Claus, P.~Kerfriden, A stable and optimally convergent latin-cutfem
  algorithm for multiple unilateral contact problems, International Journal for
  Numerical Methods in Engineering 113 (2018) 938--966.
\newblock \href {https://doi.org/https://doi.org/10.1002/nme.5694}
  {\path{doi:https://doi.org/10.1002/nme.5694}}.

\bibitem{Claus21Digital}
S.~Claus, P.~Kerfriden, F.~Moshfeghifar, S.~Darkner, K.~Erleben, C.~Wong,
  Contact modeling from images using cut finite element solvers, Advanced
  Modeling and Simulation in Engineering Sciences 8 (2021) 1--23.
\newblock \href {https://doi.org/https://doi.org/10.1186/s40323-021-00197-2}
  {\path{doi:https://doi.org/10.1186/s40323-021-00197-2}}.

\bibitem{poluektov2022cut}
M.~Poluektov, {\L}.~Figiel, A cut finite-element method for fracture and
  contact problems in large-deformation solid mechanics, Computer Methods in
  Applied Mechanics and Engineering 388 (2022) 114234.
\newblock \href {https://doi.org/https://doi.org/10.1016/j.cma.2021.114234}
  {\path{doi:https://doi.org/10.1016/j.cma.2021.114234}}.

\bibitem{JOHANSSON2019672}
A.~Johansson, B.~Kehlet, M.~G. Larson, A.~Logg, Multimesh finite element
  methods: Solving pdes on multiple intersecting meshes, Computer Methods in
  Applied Mechanics and Engineering 343 (2019) 672 -- 689.
\newblock \href {https://doi.org/https://doi.org/10.1016/j.cma.2018.09.009}
  {\path{doi:https://doi.org/10.1016/j.cma.2018.09.009}}.

\bibitem{DOKKEN2020113129}
J.~S. Dokken, A.~Johansson, A.~Massing, S.~W. Funke, A multimesh finite element
  method for the navier–stokes equations based on projection methods,
  Computer Methods in Applied Mechanics and Engineering 368 (2020) 113129.
\newblock \href {https://doi.org/https://doi.org/10.1016/j.cma.2020.113129}
  {\path{doi:https://doi.org/10.1016/j.cma.2020.113129}}.

\bibitem{JOHANSSON2020b}
A.~Johansson, M.~G. Larson, A.~Logg, Multimesh finite elements with flexible
  mesh sizes, Computer Methods in Applied Mechanics and Engineering 372 (2020)
  113420.
\newblock \href {https://doi.org/https://doi.org/10.1016/j.cma.2020.113420}
  {\path{doi:https://doi.org/10.1016/j.cma.2020.113420}}.

\bibitem{mikaeili2022concurrent}
E.~Mikaeili, S.~Claus, P.~Kerfriden, Concurrent multiscale analysis without
  meshing: Microscale representation with cutfem and micro/macro model
  blending, Computer Methods in Applied Mechanics and Engineering 393 (2022)
  114807.
\newblock \href {https://doi.org/https://doi.org/10.1016/j.cma.2022.114807}
  {\path{doi:https://doi.org/10.1016/j.cma.2022.114807}}.

\bibitem{fenics}
M.~Alnæs, J.~Blechta, J.~Hake, A.~Johansson, B.~Kehlet, A.~Logg,
  C.~Richardson, J.~Ring, M.~Rognes, G.~Wells, The fenics project version 1.5,
  Archive of Numerical Software 3 (01 2015).
\newblock \href {https://doi.org/10.11588/ans.2015.100.20553}
  {\path{doi:10.11588/ans.2015.100.20553}}.

\bibitem{Mura87}
T.~Mura, Micromechanics of Defects in Solids, Springer Netherlands, 1987.
\newblock \href {https://doi.org/10.1007/978-94-009-3489-4}
  {\path{doi:10.1007/978-94-009-3489-4}}.

\bibitem{IMANI201816489}
M.~Imani, A.~M. Goudarzi, S.~M. Rabiee, M.~Dardel, The modified mori-tanaka
  scheme for the prediction of the effective elastic properties of highly
  porous ceramics, Ceramics International 44~(14) (2018) 16489 -- 16497.
\newblock \href
  {https://doi.org/https://doi.org/10.1016/j.ceramint.2018.06.066}
  {\path{doi:https://doi.org/10.1016/j.ceramint.2018.06.066}}.

\end{thebibliography}

\end{document}